\documentclass[a4paper,11pt]{article}
\usepackage{graphics,graphicx,epstopdf}
\usepackage{amsmath,verbatim,amssymb,epsfig,color,lscape}
\usepackage{longtable}  
\usepackage{wrapfig}
\usepackage{fancyhdr} 
\usepackage{natbib}
\usepackage{lastpage}

\newcommand{\be}{\begin{equation}\begin{array}{lllllllllllllllll}}
\newcommand{\beno}{\begin{equation}\begin{array}{lllllllllllll}\nonumber}
\newcommand{\ee}{\end{array}\end{equation}}
\newcommand{\s}{\vspace{.25cm}}
\newcommand{\mR}{\mathbb{R}}

\newcommand{\mbP}{\mathbb{P}}
\newcommand{\hide}[1]{}
\usepackage{setspace}

\date{}

\RequirePackage[letterpaper, portrait, margin=1in]{geometry}

\usepackage{float}
\usepackage{standalone}

\usepackage{bm}
\usepackage{pdflscape}
\usepackage{authblk}
\usepackage[flushleft]{threeparttable}
\usepackage[citecolor=blue]{hyperref}

\usepackage{mathtools}

\newcommand{\interior}[1]{%
 {\kern0pt#1}^{\mathrm{o}}%
}
\usepackage [english]{babel}
\usepackage{algorithm,algpseudocode}
\usepackage{caption}
\usepackage[makeroom]{cancel}

\title{\textbf{Bayesian Model Selection for High-Dimensional Ising Models, With Applications to Educational Data}}
\author[1,2]{Jaewoo Park}
\author[1,2]{Ick Hoon Jin}
\author[3]{Michael Schweinberger}
\affil[1]{Department of Statistics and Data Science, Yonsei University}
\affil[2]{Department of Applied Statistics, Yonsei University}
\affil[3]{Department of Statistics, Rice University}

\begin{document}

\maketitle

\begin{abstract}
Doubly-intractable posterior distributions arise in many applications of statistics concerned with discrete and dependent data,
including physics, spatial statistics, machine learning, the social sciences, and other fields. 
A specific example is psychometrics, 
which has adapted high-dimensional Ising models from machine learning,
with a view to studying the interactions among binary item responses in educational assessments.
To estimate high-dimensional Ising models from educational assessment data,
$\ell_1$-penalized nodewise logistic regressions have been used.
Theoretical results in high-dimensional statistics show that $\ell_1$-penalized nodewise logistic regressions can recover the true interaction structure with high probability,
provided that certain assumptions are satisfied.
Those assumptions are hard to verify in practice and may be violated,
and quantifying the uncertainty about the estimated interaction structure and parameter estimators is challenging.
We propose a Bayesian approach that helps quantify the uncertainty about the interaction structure and parameters without requiring strong assumptions,
and can be applied to Ising models with thousands of parameters.
We demonstrate the advantages of the proposed Bayesian approach compared with $\ell_1$-penalized nodewise logistic regressions by simulation studies and applications to small and large educational data sets with up to 2,485 parameters.
Among other things,
the simulation studies suggest that the Bayesian approach is more robust against model misspecification due to omitted covariates than $\ell_1$-penalized nodewise logistic regressions.

\end{abstract}

\section{Introduction}
\label{intro}

Models with intractable normalizing functions arise in many applications of statistics concerned with discrete and dependent data,
including physics \citep{Ising,Ch07,GhMu20},
spatial statistics \citep{besag1974spatial,strauss1975model, goldstein2015compartmental},
machine learning \citep{RaWaLa10,Anetal12,Xuetal12,BrKa20},
and statistical network analysis \citep{robins2007introduction, hunter2012inference,caimo2013bayesian},
among others.
Models with intractable normalizing constants give rise to doubly intractable posterior distributions \citep{moller2006efficient,murray2006,Lyne15,park2018bayesian}.
As a consequence,
likelihood-based inference---whether Bayesian or non-Bayesian inference---is challenging when the likelihood function is intractable.

A specific example is the branch of psychometrics concerned with network-based approaches to learning from educational assessment data \citep{Bo08,nature14,Epskamp:2018,Marsman:2018,jin2019doubly,jeon2020}.
For example,
\citet{nature14} adapted high-dimensional Ising models from machine learning \citep{RaWaLa10,Anetal12,Xuetal12,BrKa20},
with a view to studying interactions among binary item responses in educational assessments.
To estimate high-dimensional Ising models from educational assessment data,
\citet{nature14} followed the approach of \citet{RaWaLa10} based $\ell_1$-penalized nodewise logistic regressions.
Theoretical results in high-dimensional statistics show that $\ell_1$-penalized nodewise logistic regressions can recover the true interaction structure with high probability,
provided that certain assumptions are satisfied \citep[see, e.g.,][Theorem 1, p.\ 1295]{RaWaLa10}.
Those assumptions include restricted eigenvalue and irrepresentability assumptions \citep[see assumptions (A1) and (A2) of][p.\ 1294]{RaWaLa10},
which restrict the amount of dependence among relevant predictors and the amount of dependence between relevant and irrelevant predictors \citep[see also][]{MeBu06,ZhYu06,BuGe11,Wa19,Lederer21}.
In practice,
such assumptions are hard, if not impossible to verify and may be violated.
In addition,
quantifying the uncertainty about the estimated interaction structure and parameter estimators is challenging.

We propose an alternative approach to estimating Ising models based on Bayesian variable selection methods,
which does not have such drawbacks although it does come with additional computational costs.
We combine two approaches from the Bayesian literature on doubly-intractable posterior distributions: 
(1) the double Metropolis-Hastings algorithm \citep{liang2010double} and (2) the stochastic search variable selection method \citep{George:93} using spike and slab priors \citep{Ishwaran:05}. 
We demonstrate the advantages of the proposed Bayesian approach compared with $\ell_1$-penalized nodewise logistic regressions by simulation studies and applications to small and large educational data sets with up to 2,485 parameters.
Among other things,
the simulation studies suggest that the Bayesian approach is more robust against model misspecification due to omitted covariates than $\ell_1$-penalized nodewise logistic regressions.

The remainder of our paper is organized as follows.
In Section \ref{ising}, 
we describe Ising models along with computational and statistical challenges arising from Ising models,
and existing approaches for addressing them.
In Section \ref{mcmc}, 
we propose a Bayesian algorithm for estimating Ising models.
In Section \ref{sim}, 
we assess the performance of the method by using simulated data.
In Section \ref{app}, 
we apply the Bayesian approach to three educational data sets. 

\section{Statistical framework}
\label{ising}

We introduce the statistical framework used throughout the remainder of our paper.
We first review Ising models along with generalizations (Section \ref{ising2}).
We then discuss computational and statistical challenges arising from Ising models (Section \ref{comp}),
along with non-Bayesian and Bayesian approaches for addressing them (Sections \ref{nonbayes} and \ref{bayes}).
 
\subsection{Ising models and generalizations}
\label{ising2} 

We introduce Ising models along with generalizations,
with a view to studying interactions among binary item responses.

We consider binary item response data $\mathbf{X} \in \{0, 1\}^{n \times p}$ consisting of item responses $X_{i,j} \in \{0, 1\}$,
where $X_{i,j} = 1$ indicates that the response of the $i$-th respondent to the $j$-th item is correct and $X_{i,j} = 0$ otherwise ($i = 1, \dots, n$,\; $j = 1, \dots, p$). 
A natural approach to studying interactions among binary responses is based on Ising models \citep{nature14}.
Ising models---as used in psychometrics, statistics, and machine learning---are undirected graphical models for binary responses \citep{La96,graphical.models},
represented in exponential-family form \citep{Su19}.
The probability mass function of Ising models is of the form
\be
\mbP_{\bm{\theta}}(\mathbf{X} = \mathbf{x})
&=& \displaystyle\prod_{i=1}^n\, \displaystyle\frac{1}{\kappa(\boldsymbol{\theta})}\, \exp\left(\sum_{j=1}^{p}\, \beta_{j}\, x_{i,j} + \sum_{j<k}^p\, \gamma_{j,k}\, x_{i,j}\, x_{i,k}\right), 
& \boldsymbol{\theta}=(\boldsymbol{\beta},\, \boldsymbol{\gamma})\, \in\, \mathbb{R}^{p + \binom{p}{2}}.
\label{ergm1}
\ee
Here,
$\beta_j \in \mR$ can be interpreted as the easiness of item $j$ and $\gamma_{j,k} \in \mR$ can be interpreted as the weight of a pairwise interaction of two distinct items $j$ and $k$.
If $\gamma_{j,k} = 0$,
responses to items $j$ and $k$ are independent conditional on all other item responses.
The conditional independencies implied by the Ising model can be represented by a conditional independence graph,
as in other undirected graphical models \citep{graphical.models}.
In other words,
each item $j$ is represented by a node in an undirected graph,
and two distinct items $j$ and $k$ are not connected by an edge when $\gamma_{j,k} = 0$,
that is,
when responses to items $j$ and $k$ are independent conditional on all other item responses.
Otherwise items $j$ and $k$ are connected by an edge.
Examples of conditional independence graphs can be found in Sections \ref{sim} and \ref{app}.
Generalizations of Ising models from second-order interactions of the form $x_{i,j}\, x_{i,k}$ to third-order interactions of the form $x_{i,j}\, x_{i,k}\, x_{i,l}$ and higher-order interactions are possible:
see,
e.g.,
the seminal works of \citet{Bj74} and \citet{FoSd86} on related models in spatial statistics and statistical network analysis,
respectively.
Those generalizations share the same exponential-family platform as Ising models and therefore pose similar computational and statistical challenges,
although computing statistics involving third- and higher-order interactions can increase the computational burden.
For the sake of concreteness,
we focus here on Ising models with pairwise interactions.

It is worth noting that there is a useful relationship between the Ising model on the one hand and logistic regression on the other hand,
because the log odds of the conditional probability of $X_{i,j}=1$,
given all other item responses,
is of the form
\be
\label{logit0}
\log \dfrac{\mbP_{\boldsymbol{\theta}}(X_{i,j} = 1 \mid X_{i,k},\, k \neq j)}{1 - \mbP_{\boldsymbol{\theta}}(X_{i,j} = 1 \mid X_{i,k},\, k \neq j)}
&=& \beta_j + \displaystyle\sum_{k=1:\, k \neq j}^p \gamma_{j,k}\, x_{i,k}.
\ee
The relationship between the Ising model and logistic regression has at least two advantages.
First,
it shows that the parameters $\beta_j$ and $\gamma_{j,k}$ can be interpreted in terms of conditional log odds and log odds ratios \citep[e.g.,][]{Ag02,StScBoMo18}.
Second,
it suggests that the conditional independence graph of the Ising model along with its parameters can be estimated by using $\ell_1$-penalized nodewise logistic regressions \citep{RaWaLa10,nature14}.
We describe the $\ell_1$-penalized nodewise logistic regression approach of \citet{RaWaLa10} and \citet{nature14} in Section \ref{nonbayes}.
Having said that,
it is important to keep in mind that the Ising model is more than logistic regression.
While conventional logistic regression models assume that item responses are independent,
the Ising model induces dependence among item responses.
Indeed,
according to \eqref{logit0},
$\gamma_{j,k} \neq 0$ implies that the log odds of the conditional probability of $X_{i,j}=1$ depends on the value $x_{i,k}$ of $X_{i,k}$,
while the log odds of the conditional probability of $X_{i,k}=1$ depends on the value $x_{i,j}$ of $X_{i,j}$,
and so forth ($j \neq k = 1, \dots, p$,\; $i = 1, \dots, n$).

\subsection{Computational and statistical challenges}
\label{comp}

While useful for studying interactions among binary responses,
Ising models give rise to formidable computational and statistical challenges.
Chief among them is the fact that the normalizing function $\kappa(\boldsymbol{\theta})$ is a sum over all $2^{p} = \exp(p\, \log 2)$ possible combinations of item responses,
which cannot be computed by complete enumeration of all $\exp(p\, \log 2)$ possible combinations of item responses unless $p$ is small (e.g., $p \ll 20$).

There are two scenarios in which the normalizing function simplifies and can be computed in reasonable time (polynomial in $p$).
First,
if the interaction weights $\gamma_{j,k}$ of all distinct pairs of items $j$ and $k$ are zero,
the conditional independence graph is empty in the sense that it contains no edges,
the item responses are independent Bernoulli random variables,
and the normalizing function simplifies.
Second,
if the item responses are dependent but the conditional independence graph of the Ising model is decomposable in the graph-theoretic sense of the word and all cliques and separators of the conditional independence graph are small,
the normalizing function likewise simplifies \citep{La96,Wh09}.

Having said that,
the assumptions that the conditional independence graph is empty (first scenario) or decomposable (second scenario) are restrictive,
because both assumptions limit the interactions among item responses that can be captured by Ising models.
Aside from these two scenarios of limited interest,
it is challenging to compute the likelihood function in reasonable time unless $p$ is small (e.g., $p \ll 20$).

\subsection{Non-Bayesian approaches}
\label{nonbayes}

A well-known non-Bayesian approach to estimating Ising models is the eLasso approach of \citet{nature14},
which sidesteps computations of the normalizing function $\kappa(\bm\theta)$ and serves as a benchmark throughout the remainder of our paper.

The eLasso approach of \citet{nature14} is based on the $\ell_1$-penalized nodewise logistic regression approach of \citet{RaWaLa10}.
The main idea is that, 
while the joint probability distribution \eqref{ergm1} of all item responses may be intractable,
the full conditional probability distributions of item responses are Bernoulli distributions and are hence tractable.
In fact,
the log odds of the conditional probability of $X_{i,j}=1$, given all other item responses, is of the form \eqref{logit0},
which resembles a logistic regression of $X_{i,j}$ on all other item responses.
\citet{RaWaLa10} therefore suggested to learn the conditional independence graph of the Ising model along with its parameters by learning the neighborhoods of the items in the conditional independence graph through logistic regressions of item responses $X_{i,j}$ on all other item responses.
To encourage the neighborhoods of items in the conditional independence graph to be sparse,
the nodewise logistic regressions are subject to $\ell_1$-penalties.
The regularization parameter of the $\ell_1$-penalties is based on the Extended Bayesian Information Criterion (EBIC) of \citet{ChCh08}.
% The computational burden is $O(\max\{p,\, n\}\, p^3)$.
The conditional independence graph is estimated by combining the estimates of the neighborhoods obtained from the $\ell_1$-penalized nodewise logistic regressions by using the so-called AND rule of \citet{MeBu06},
that is,
the estimated conditional independence graph contains an edge between two distinct items $i$ and $j$ if and only if the estimated neighborhoods of $i$ and $j$ both contain an edge between $i$ and $j$.
More details can be found in \citet{nature14}.

A computational advantage of eLasso is that $\ell_1$-penalized nodewise logistic regressions do not involve intractable normalizing functions and can be computed separately.
The fact that the $\ell_1$-penalized nodewise logistic regressions can be computed separately opens the door to parallel computing on multicore computers or computing clusters,
facilitating large-scale computing.
The theoretical properties of the $\ell_1$-penalized nodewise logistic regression approach have been studied by \citet{RaWaLa10}.
Other theoretical work on Ising models along with generalizations can be found in \citet{Anetal12}, \citet{Xuetal12}, and \citet{BrKa20}.

While attractive on computational grounds,
there is no such thing as a free lunch:
The approach of \citet{RaWaLa10} and \citet{nature14} can recover the true conditional independence graph with high probability,
provided that strong assumptions are satisfied.
Those assumptions are hard, if not impossible to verify and may be violated in practice,
as discussed in Section \ref{intro}.
Worse,
quantifying the uncertainty about the conditional independence graph and parameter estimators is challenging.
We therefore adopt a Bayesian approach,
which comes at additional computational costs but helps capture the uncertainty about the interactions among items and incorporate prior information (provided prior information is available).

\subsection{Bayesian approaches}
\label{bayes}

We review Bayesian alternatives to eLasso.
Bayesian approaches to Ising models face the same obstacles as non-Bayesian approaches \citep{RaWaLa10,nature14}:
the normalizing function $\kappa(\boldsymbol{\theta})$,
which is infeasible to compute in reasonable time except in special cases of limited interest (described in Section \ref{comp});
and the fact that the number of parameters $p + \binom{p}{2}$ is large even when $p$ is not large:
e.g., 
in the application to the Korean middle school data in Section \ref{sec:data3},
we have $p=70$ items and $p + \binom{p}{2} = $ 2,485 parameters.
As a result,
in many scenarios the likelihood function is intractable and the posterior distribution is doubly intractable.

In the Bayesian literature,
a popular approach to sidestepping computations of intractable likelihood functions is Approximate Bayesian Computation (ABC) \citep{Preetal99,beaumont2002approximate, Maetal03, Sisson1760, Toni09, marin2012approximate, YiBu20}.
A simple form of ABC samples a candidate parameter vector from a proposal distribution (e.g., a multivariate Normal distribution) and,
given the candidate parameter vector, 
simulates a synthetic sample from probability mass function \eqref{ergm1} using the candidate parameter vector.
If the observed and simulated data are close in a well-defined sense,
the candidate parameter vector is accepted,
otherwise it is rejected.
To assess whether the observed and simulated data are close,
some summary statistics are chosen to compare the observed and simulated data (e.g., sufficient statistics),
along with a distance function that compares the summary statistics of the observed and simulated data (e.g., the Euclidean distance between the summary statistics of the observed and simulated data).
The accepted parameter vectors are then regarded as a sample from an approximation to the posterior distribution.
More sophisticated versions of ABC exist \citep[e.g.,][]{Maetal03,Sisson1760, Toni09, marin2012approximate, YiBu20}.
That said,
questions have been raised about the usefulness of ABC in Bayesian model selection problems for Gibbs random fields \citep{Robert15112}.
Since Gibbs random fields are closely related to Ising models by virtue of being exponential-family models \citep{Su19}, 
it is not clear whether ABC would be useful for learning the conditional independence graph of Ising models.

Other Bayesian approaches have been developed for models with intractable normalizing functions. 
Many of them fall into one of two broad categories: 
(1) likelihood approximation approaches, 
which approximate the normalizing function $\kappa(\boldsymbol{\theta})$ by Markov chain Monte Carlo and plug the approximation of $\kappa(\boldsymbol{\theta})$ into the acceptance probability of Metropolis-Hastings algorithms \citep{Ko04,atchade2006adaptive, liang2013monte, Lyne15, alquier2014noisy, park2019function}; 
and (2) auxiliary variable approaches, 
which augment the posterior distributions by auxiliary variables so that the normalizing function $\kappa(\boldsymbol{\theta})$ in the acceptance probability of Metropolis-Hastings algorithms cancels \citep{moller2006efficient, murray2006, liang2010double, liang2015adaptive}. 
A full-fledged discussion of these approaches is beyond the scope of our paper,
but we highlight some of them and refer to \citet{park2018bayesian} for a review of other approaches.
For example,
the auxiliary-variable methods of \citet{moller2006efficient} and \citet{murray2006} rely on perfect sampling of auxiliary variables \citep[e.g.,][]{propp1996exact,Bu12}.
However,
perfect sampling can be expensive in terms of computing time.
To address these computational issues, 
\citet{liang2010double} developed a double Metropolis-Hastings  algorithm by generating an auxiliary variable with a finite number of Metropolis-Hastings updates. 
Despite sampling from an approximation of the target distribution rather than the target distribution itself \citep{park2018bayesian},
the double Metropolis-Hastings algorithm is the most practical approach on computational grounds.
Many of the aforementioned approaches have been explored in the popular class of exponential-family random graph models \citep[ERGMs,][]{ergm.book,HuKrSc12,ScKrBu17},
which share the exponential-family platform with Ising models and Gibbs random fields and therefore give rise to similar computational issues:
see,
e.g.,
\cite{KoRoPa09}, \citet{caimo2011bayesian}, \citet{AtLaRo12}, \citet{bergm.jss}, \citet{caimo2013bayesian}, \citet{JiYuLi13}, and \citet{CaGo20m} for ERGMs and \citet{Ev12}, \citet{ScHa13}, and \citet{thiemichen2016bayesian} for ERGMs with latent structure.
Scalable approaches were developed by \cite{bouranis2017efficient} and \cite{bouranis2018bayesian},
based on pseudolikelihood functions.
While promising,
these approaches have been applied in low-dimensional settings with fewer than $10$ parameters,
not in high-dimensional settings with thousands of parameters.
As a consequence,
we focus here on a Bayesian approach based on a double Metropolis-Hastings algorithm,
motivated by its feasibility in high-dimensional settings.

\section{Bayesian algorithm}
\label{mcmc}

We propose a Bayesian algorithm based on the stochastic search variable selection approach \citep{George:93}, 
and incorporate a spike and slab prior \citep{Ishwaran:05} into a double Metropolis-Hastings algorithm \citep{liang2010double}. 
We introduce spike and slab priors in Section \ref{priors},
and then introduce a spike and slab double Metropolis-Hastings algorithm in Section \ref{db}.

\subsection{Spike and slab priors}
\label{priors}

To estimate the conditional independence graph of the Ising model,
we assume that the coordinates $\theta_i$ of the parameter vector $\bm{\theta} \in \mR^{p + \binom{p}{2}}$ are generated from a spike and slab prior of the form
\be
\label{ssprior}
\theta_i \mid \lambda_i,\, \sigma^2,\, \omega & \overset{\text{ind}}{\sim} & \lambda_i\, N(0,\, \omega^2\; \sigma^2) + (1-\lambda_i)\, N(0,\, \sigma^2)\s
\\
\lambda_i  &\overset{\text{iid}}{\sim}& \mbox{Bernoulli}(1/2)\s
\\
\dfrac{1}{\sigma^2} &\sim& \mbox{Uniform}(4,\, 100)\s
\\ 
\omega &\sim& 1 + Y,\hspace{2cm} Y \sim \mbox{Gamma}(1,\, 1/100).
\ee
The indicator $\lambda_i \in \{0, 1\}$ determines whether the parameter $\theta_i \in \mR$ is generated from the Normal distribution $N(0,\, \omega^2\, \sigma^2)$ with mean $0$ and variance $\omega^2\, \sigma^2 > 0$ ($\lambda_i=1$) or from the Normal distribution $N(0,\, \sigma^2)$ with mean $0$ and variance $\sigma^2 > 0$ ($\lambda_i=0$).
We assume that the prior of $\lambda_i$ is Bernoulli$(1/2)$,
that is,
the prior probabilities of the events $\{\lambda_i = 1\}$ and $\{\lambda_i = 0\}$ are both $1/2$. 
The parameter $\sigma^2 > 0$ controls the variance of the distribution $N(0,\, \sigma^2)$.
The Uniform$(4,\, 100)$ prior of the inverse variance $1/\sigma^2$ ensures that the prior of the variance $\sigma^2$ is a right-skewed distribution taking values in the range $(1/100,\, 1/4)$,
with a mean of approximately $1/30$ and a long upper tail stretching from $1/30$ to $1/4$.
The distribution $N(0,\, \sigma^2)$ is called a spike distribution,
because the small variance $\sigma^2$ implies that the bulk of the probability mass of $N(0,\, \sigma^2)$ is concentrated in a small neighborhood of $0$ and therefore the distribution resembles a spike at $0$.
The parameter $\omega > 1$ ensures that the variance $\omega^2\, \sigma^2$ of the distribution $N(0,\, \omega^2\, \sigma^2)$ exceeds the variance $\sigma^2$ of the spike distribution $N(0,\, \sigma^2)$.
The expectation of $Y \sim$ Gamma$(1,\, 1/100)$ is 100,
which implies that the expectation of $\omega = 1 + Y$ is 101.
In other words,
the standard deviation $\omega\, \sigma$ of the distribution $N(0,\, \omega^2\; \sigma^2)$ is expected to be approximately 100 times larger than the standard deviation $\sigma$ of the spike distribution $N(0,\, \sigma^2)$.
As a consequence,
the distribution $N(0,\, \omega^2\; \sigma^2)$ is much flatter than the spike distribution $N(0,\, \sigma^2)$ and is called a slab distribution.

To introduce the posterior distribution,
write $q = p + \binom{p}{2}$,\,
$\bm\theta = (\theta_1, \dots, \theta_q)$,\,
and $\bm\lambda = (\lambda_1, \dots, \lambda_q)$.
According to \eqref{ssprior},
the prior probability density function of $\bm\theta$, $\bm\lambda$, $\sigma^2$, and $\omega$ is of the form
\beno
\pi(\bm\theta,\; \bm\lambda,\; \sigma^2,\; \omega)
&=& \pi(\bm\theta \mid \bm\lambda,\, \sigma^2,\, \omega)\; \pi(\bm\lambda)\; \pi(\sigma^2)\; \pi(\omega),
\ee
where the conditional and marginal prior probability density functions $\pi(\boldsymbol{\theta} \mid \bm\lambda,\, \sigma^2,\, \omega)$, $\pi(\bm\lambda)$, $\pi(\sigma^2)$, and $\pi(\omega)$ follow from \eqref{ssprior};
note that $\pi(\bm\lambda)$ is a probability density function with respect to counting measure,
whereas the others are probability density functions with respect to Lebesgue measure.
We refer to all of them as probability density functions,
with the tacit understanding that these probability density functions are taken with respect to a dominating measure with suitable support \citep{shao}.
The posterior probability density function of $\bm{\theta}$,
$\bm\lambda$,
$\sigma^2$,
and $\omega$,
given observed data $\bm{x}$,
is then proportional to
\be
\label{posterior}
\pi(\boldsymbol{\theta},\, \boldsymbol{\lambda},\, \sigma^2,\, \omega \mid \mathbf{x}) 
&\propto& \mbP_{\bm\theta}(\mathbf{X} = \mathbf{x})\; \pi(\boldsymbol{\theta} \mid \boldsymbol{\lambda},\, \sigma^2,\, \omega)\; \pi(\boldsymbol{\lambda})\; \pi(\sigma^2)\; \pi(\omega),
\ee
where the probability mass function $\mbP_{\bm\theta}(\mathbf{X} = \mathbf{x})$ is of the form \eqref{ergm1}.

\subsection{Spike and slab double Metropolis-Hastings algorithm}
\label{db}

The posterior distribution \eqref{posterior} involves the normalizing function of $\mbP_{\bm\theta}(\mathbf{X} = \mathbf{x})$,
which is intractable except in special cases of limited interest (described in Section \ref{comp}).
We therefore propose a spike and slab double Metropolis-Hastings algorithm for approximating the posterior distribution.

A description of a spike and slab double Metropolis-Hastings algorithm can be found in Algorithm \ref{spikedouble Metropolis-Hastings algorithmalg} on page \pageref{algorithm1}.
Consider the parameters at the $t$-th iteration,
denoted by
\beno
(\boldsymbol{\theta}^{(t)},\, \boldsymbol{\lambda}^{(t)},\, \sigma^{2(t)},\, \omega^{(t)}) 
&=& (\theta_1^{(t)},\, \dots,\, \theta_q^{(t)},\, \lambda_1^{(t)},\, \dots,\, \lambda_q^{(t)},\, \sigma^{2(t)},\, \omega^{(t)}).
\ee
At iteration $t=0$,
we initialize the parameters by setting $\boldsymbol{\lambda}^{(0)}$ = $(1, \dots, 1)$ and sampling\linebreak 
$\boldsymbol{\theta}^{(0)}$ $\sim$ $\mbox{Uniform}(-5,\,  5)$,
$1\, /\, \sigma^{2(0)}$ $\sim$ $\mbox{Uniform}(4,\, 100)$,
and $\omega^{(0)} = 1 + Y$,
where $Y \sim \mbox{Gamma}(1,\, 1/100)$. 
At iteration $t+1$,
the parameters are updated by cycling through the parameters as follows.

{\bf Parameters $\theta_i^{(t)}$ ($i = 1, \dots, q$).}
The parameters $\theta_i^{(t)}$ can be updated by sampling from the conditional distributions
\be
\label{stage1}    
\theta_i^{(t+1)} 
\mid \mathbf{x},\; \boldsymbol{\theta}_{-i}^{(t)},\; \lambda_{i}^{(t)},\; \sigma^{2(t)},\; \omega^{(t)},
\ee
where $\boldsymbol{\theta}_{-i}^{(t)} = (\theta_{1}^{(t+1)},\, \dots,\, \theta_{i-1}^{(t+1)},\, \theta_{i+1}^{(t)},\, \dots,\, \theta_{q}^{(t)} )$. 
In principle,
we could update the parameters $\theta_i^{(t)}$ by using Metropolis-Hastings algorithms.
However,
the acceptance probability of conventional Metropolis-Hastings algorithms involves ratios of $\mbP_{\bm\theta}(\mathbf{X} = \mathbf{x})$,
whose normalizing functions cannot be computed (leaving aside special cases of limited interest, described in Section \ref{comp}).
As a consequence,
computing the acceptance probability of conventional Metropolis-Hastings algorithms is not feasible when the normalizing function cannot be computed.
We therefore take advantage of a double Metropolis-Hastings algorithm \citep{liang2010double},
which sidesteps computations of the normalizing functions by cleverly augmenting the posterior distribution with auxiliary variables.
The basic idea is to augment the posterior distribution by an auxiliary variable $\bm{Y} \in \{0, 1\}^{n \times p}$ with probability mass function \eqref{ergm1}.
A double Metropolis-Hastings algorithm then samples from the augmented posterior distribution as follows.
First,
it generates a candidate parameter $\theta'_{i}$ from a Normal distribution centered at the current value $\theta_i^{(t)}$ of parameter $\theta_i$,
where the standard deviation of the Normal distribution is chosen so that the acceptance rate of the Metropolis-Hastings algorithm is between $2/10$ and $3/10$.
Given the candidate parameter  $\theta'_{i}$,
it then generates an auxiliary variable $\bm{Y} \in \{0, 1\}^{n \times p}$ with probability mass function \eqref{ergm1} and parameter vector $(\theta'_{i},\, \boldsymbol{\theta}_{-i}^{(t)})$ using $m = 10\, n$ Metropolis-Hastings steps.
%\cite{liang2010double} and \citet{park2018bayesian} recommend to choose $m$ proportional to $n$. 
%We therefore use $m=10\, n$ Metropolis-Hastings steps to generate the auxiliary variable $\bm{Y}$.
The double Metropolis-Hastings algorithm then accepts the candidate parameter $\theta'_{i}$ with probability $\varphi_1 = \min\{1,\, \rho_1\}$,
where $\rho_1$ is given by \eqref{mhratio}.
The resulting acceptance probability of the double Metropolis-Hastings algorithm does not involve normalizing functions,
because all normalizing functions cancel.
%To detect whether the choice of $m = 10\, n$ is large enough,
%\citet{park2018bayesian} recommend to conduct a second run of the double Metropolis-Hastings algorithm with an increased $m$,
%e.g.,
%$m=20\, n$,
%with a view to detecting whether there are substantial changes in the approximation of the posterior distribution as $m$ is increased.
%We follow \citet{park2018bayesian}'s advice in the applications in Section \ref{app}.

{\bf Indicators $\lambda_{i}^{(t)}$ ($i = 1, \dots, q$).}
The indicators $\lambda_{i}^{(t)}$ can be updated by sampling from the conditional distributions
\be
\label{stage2}
\lambda_i^{(t+1)} \mid \mathbf{x},\; \theta_i^{(t+1)},\; \boldsymbol{\lambda}_{-i}^{(t)},\; \sigma^{2(t)},\; \omega^{(t)},
\ee
where $\boldsymbol{\lambda}_{-i}^{(t)} = (\lambda_{1}^{(t+1)},\, \cdots,\, \lambda_{i-1}^{(t+1)},\, \lambda_{i+1}^{(t)},\, \cdots,\, \lambda_{q}^{(t)})$. 
The conditional distribution in \eqref{stage2} is a Bernoulli distribution.
The parameter of the Bernoulli distribution is stated in the description of Algorithm \ref{spikedouble Metropolis-Hastings algorithmalg} on page \pageref{algorithm1}.
%By cycling through \eqref{stage1} and \eqref{stage2}, 
%we can obtain updates $\boldsymbol{\theta}^{(t+1)}$ and $\boldsymbol{\lambda}^{(t+1)}$. 

{\bf Parameters $\sigma^{2(t)}$ and $\omega^{(t)}$.}
The parameters $\sigma^{2(t)}$ and $\omega^{(t)}$ can be updated by Metropolis-Hastings algorithms that sample from the conditional distributions
\beno
\sigma^{2(t+1)} \mid \mathbf{x},\; \boldsymbol{\theta}^{(t+1)},\; \boldsymbol{\lambda}^{(t+1)},\; \omega^{(t)}
\ee
% &=& \pi(\boldsymbol{\theta}^{(t+1)} \mid \boldsymbol{\lambda}^{(t+1)},\, \sigma^{2(t)},\, \omega^{(t)})\; \pi(\sigma^{2(t)})\s
and
\beno
\omega^{(t+1)} \mid \mathbf{x},\; \boldsymbol{\theta}^{(t+1)},\; \boldsymbol{\lambda}^{(t+1)},\; \sigma^{2(t+1)},
%&=& \pi(\boldsymbol{\theta}^{(t+1)} \mid \boldsymbol{\lambda}^{(t+1)},\, \sigma^{2(t+1)},\, \omega^{(t)})\; \pi(\omega^{(t)}).
\ee
respectively.
More details can be found in the description of Algorithm \ref{spikedouble Metropolis-Hastings algorithmalg} on page \pageref{algorithm1}.
\begin{algorithm}[htbp]
\label{algorithm1}
\caption{Spike and slab double Metropolis-Hastings algorithm.
As proposal distributions $w(. \mid .)$,
we use independent Normal distributions centered at the current values of the parameters,
with standard deviations set to achieve an acceptance rate between $2/10$ and $3/10$.}
\label{spikedouble Metropolis-Hastings algorithmalg}
\begin{algorithmic}
\normalsize

\s

\State At iteration $t+1$:
\\

\State \textbf{Part I: Given $\boldsymbol{\theta}^{(t)},\, \boldsymbol{\lambda}^{(t)},\, \sigma^{2(t)}$, and $\omega^{(t)}$, 
update ${\theta}_i^{(t)}$ and ${\lambda}_i^{(t)}$ as follows ($i = 1, \dots, q$).}\\

\State {\it{Step 1.}} Propose $\theta'_{i} \sim w(\cdot\mid\theta^{(t)}_{i})$.\s
\\
 
\State {\it{Step 2.}} Generate an auxiliary variable $\bm{Y} \in \{0, 1\}^{n \times p}$ from probability mass function \eqref{ergm1} with parameter vector $(\theta'_{i},\, \boldsymbol{\theta}_{-i}^{(t)})$ by using $m = 10\, n$ Metropolis-Hastings steps.\s
%: $\mathbf{y} \sim f(\mathbf{x} \mid \theta'_{i},\, \boldsymbol{\theta}_{-i}^{(t)})$, 
%where $\boldsymbol{\theta}_{-i}^{(t)} = (\theta_{1}^{(t+1)},\, \cdots,\, \theta_{i-1}^{(t+1)},\, \theta_{i+1}^{(t)},\, \cdots,\, \theta_{q}^{(t)})$.\s
\\

\State {\it{Step 3.}} Set $\theta^{(t+1)}_{i}=\theta'_{i}$ with probability $\varphi_1 = \min\{1,\, \rho_1\}$,
where
\be
\label{mhratio}
\rho_1 = \dfrac{\mbP_{\theta'_{i},\, \boldsymbol{\theta}_{-i}^{(t)}}(\mathbf{X} = \mathbf{x})\; \mbP_{\theta^{(t)}_{i},\, \boldsymbol{\theta}_{-i}^{(t)}}(\mathbf{Y} = \mathbf{y})\; \pi(\theta'_i \mid \lambda^{(t)}_i,\, \sigma^{2(t)},\, \omega^{(t)})\; w(\theta^{(t)}_{i} \mid \theta'_{i})}{\mbP_{\theta_{i}^{(t)},\, \boldsymbol{\theta}_{-i}^{(t)}}(\mathbf{X} = \mathbf{x})\; \mbP_{\theta'_{i},\, \boldsymbol{\theta}_{-i}^{(t)}}(\mathbf{Y} = \mathbf{y})\; \pi(\theta^{(t)}_i \mid \lambda^{(t)}_i,\, \sigma^{2(t)},\, \omega^{(t)})\; w(\theta'_{i} \mid \theta^{(t)}_{i})},
\ee
otherwise set $\theta^{(t+1)}_{i} = \theta^{(t)}_{i}$.\s
\\

\State {\it{Step 4.}} 
Set $\lambda^{(t+1)}_{i} = 1$ with probability $a / (a+b)$ and otherwise set $\lambda^{(t+1)}_{i} = 0$,
where 
\beno
\label{a.b}
a &=& \pi(\theta_{i}^{(t+1)} \mid \lambda_i^{(t+1)} = 1,\, \boldsymbol{\lambda}_{-i}^{(t)},\, \sigma^{2(t)},\, \omega^{(t)})\; \pi(\boldsymbol{\lambda}_{-i}^{(t)},\; \lambda_i^{(t+1)} = 1)\s\s
\\
b &=& \pi(\theta_{i}^{(t+1)} \mid \lambda_i^{(t+1)} = 0,\, \boldsymbol{\lambda}_{-i}^{(t)},\, \sigma^{2(t)},\, \omega^{(t)})\; \pi(\boldsymbol{\lambda}_{-i}^{(t)},\; \lambda_i^{(t+1)} = 0).
\ee

\s

\\

\State \textbf{Part II: Given $\boldsymbol{\theta}^{(t+1)},\, \boldsymbol{\lambda}^{(t+1)},\, \sigma^{2(t)}$, and $\omega^{(t)}$,
update $\sigma^{2(t)}$ and $\omega^{(t)}$.}\s
\\

\State {\it{Step 5.}} Propose $\sigma^{2 \prime} \sim w(\cdot\mid\sigma^{2(t)})$ and accept it with probability $\varphi_2 = \min\{1,\, \rho_2\}$,
where 
$$\rho_2 = \dfrac{\pi(\boldsymbol{\theta}^{(t+1)} \mid \boldsymbol{\lambda}^{(t+1)},\, \sigma^{2 \prime},\, \omega^{(t)})\; \pi(\sigma^{2 \prime})\; w(\sigma^{2(t)} \mid \sigma^{2 \prime})}{\pi(\boldsymbol{\theta}^{(t+1)} \mid \boldsymbol{\lambda}^{(t+1)},\, \sigma^{2(t)},\, \omega^{(t)})\; \pi(\sigma^{2(t)})\; w(\sigma^{2 \prime} \mid \sigma^{2(t)})},$$
otherwise set $\sigma^{2(t+1)}=\sigma^{2(t)}$.\s
\\

\State {\it{Step 6.}} Propose $\omega' \sim w(\cdot\mid\omega^{(t)})$ and accept it with probability $\varphi_3 = \min\{1,\, \rho_3\}$,
where
$$\rho_3 = \dfrac{\pi(\boldsymbol{\theta}^{(t+1)} \mid \boldsymbol{\lambda}^{(t+1)},\, \sigma^{2(t+1)},\, \omega')\; \pi(\omega')\; w(\omega^{(t)}\mid\omega')}{\pi(\boldsymbol{\theta}^{(t+1)} \mid \boldsymbol{\lambda}^{(t+1)},\, \sigma^{2(t+1)},\, \omega^{(t)})\; \pi(\omega^{(t)})\; w(\omega' \mid \omega^{(t)})},$$
otherwise set $\omega^{(t+1)} = \omega^{(t)}$.
\end{algorithmic}
\end{algorithm}

{\bf Posterior estimates of edges and parameters.}
The edges in the conditional independence graph of the Ising model and its parameters can be estimated based on a Markov chain Monte Carlo sample as follows.
Consider two distinct items $j$ and $k$.
We first estimate the posterior interaction probability of the event that the indicator $\lambda_i$ corresponding to the interaction weight $\gamma_{j,k}$ equals $1$ by the corresponding Markov chain Monte Carlo sample proportion.
If the posterior interaction probability is at least $1/2$,
we connect items $j$ and $k$ by an edge in the conditional independence graph,
otherwise we do not connect them.
If items $j$ and $k$ are connected by an edge,
the interaction weight $\gamma_{j,k}$ is estimated by its posterior mean,
otherwise $\gamma_{j,k}$ is estimated as $0$.
The intercepts $\beta_j$ are estimated by an analogous procedure,
by first determining whether $\beta_j$ is non-zero and then estimating $\beta_j$ by its posterior mean,
provided it is non-zero.

\subsection{Implementation details}
\label{sec:impl}

The Bayesian approach is implemented in {\tt C++} and {\tt R} using packages {\tt Rcpp} and {\tt RcppArmadillo} \citep{eddelbuettel2011rcpp},
while eLasso is implemented in {\tt R} package {\tt IsingFit} \citep{isingfit}.
We used {\tt R} version 3.6.1,
{\tt RcppArmadillo} version 0.9.600.4.0,
and {\tt IsingFit} version 0.3.1.
Implementation details on eLasso can be found in Section \ref{nonbayes} and \citet{nature14}.
We use the same settings as \citet{nature14},
which are the default settings in {\tt R} package {\tt IsingFit} version 0.3.1 \citep{isingfit}.
Implementation details on the Bayesian algorithm can be found in Section \ref{db}.
We check the convergence of the Bayesian algorithm by computing the Monte Carlo standard errors (MCSE) calculated by batch means \citep{jones2006fixed, flegal2008markov}.
The Bayesian algorithm is run until the MCSE is at most 3/100. 
All algorithms were run on dual 32 core AMD Ryzen Threadripper 2990WX processors. 
The source code can be downloaded from https://github.com/jwpark88/itemBayes. 

\section{Simulation studies}
\label{sim}

We compare the Bayesian approach with eLasso by conducting two simulation studies.
In both simulation studies,
we are interested in comparing how well the Bayesian approach can recover the conditional independence graph of the Ising model compared with eLasso.
The first of the two simulation studies estimates the conditional independence graph under a correct model specification,
whereas the second simulation study estimates the conditional independence graph under an incorrect model specification.
To conduct simulation studies under correct as well as incorrect model specifications,
we introduce a generalized Ising model with covariates.
The probability mass function of a generalized Ising model with covariate vector $\mathbf{c} \in \{0, 1\}^{n}$ is of the form
\be
\label{ex}
\mbP_{\boldsymbol{\theta}}(\mathbf{X} = \mathbf{x})
&=& \displaystyle\prod_{i=1}^n\; \displaystyle\frac{1}{\kappa(\boldsymbol{\theta})}\; \exp\left(\displaystyle\sum_{j=1}^p \alpha\, c_{i}\, x_{i,j} + \sum_{j=1}^{p}\, \beta_{j}\, x_{i,j} + \sum_{j<k}^p\, \gamma_{j,k}\, x_{i,j}\, x_{i,k}\right)\s
\\
&=& \displaystyle\prod_{i=1}^n\; \displaystyle\frac{1}{\kappa(\boldsymbol{\theta})}\; \exp\left(\displaystyle\sum_{j=1}^p (\alpha\, c_{i} + \beta_{j})\, x_{i,j} + \sum_{j<k}^p\, \gamma_{j,k}\, x_{i,j}\, x_{i,k}\right),
\ee
where the covariate vector $\mathbf{c}$ is assumed to be non-constant,
that is,
the covariates of respondents are neither all $0$ nor all $1$.
Here,
$\alpha \in \mR$ is the weight of the covariate term and $\boldsymbol{\theta}=(\alpha,\, \boldsymbol{\beta},\, \boldsymbol{\gamma})\, \in\, \mathbb{R}^{p + \binom{p}{2}+1}$.
If $\alpha=0$,
the generalized Ising model \eqref{ex} reduces to the Ising model \eqref{ergm1},
otherwise the generalized Ising model \eqref{ex} can be viewed as a generalization of the Ising model \eqref{ergm1} with a covariate term.
The covariate term is a weighted sum of indicators $c_i \in \{0, 1\}$,
e.g.,
in applications to educational data the indicators might be indicators of whether respondents $i$ are female.
To gain insight into the effect of the covariate term on the item responses,
it is instructive to inspect the log odds of the conditional probability of $X_{i,j}=1$, 
given all other item responses:
\be
\label{logit}
\log \dfrac{\mbP_{\boldsymbol{\theta}}(X_{i,j} = 1 \mid X_{i,k},\, k \neq j)}{1 - \mbP_{\boldsymbol{\theta}}(X_{i,j} = 1 \mid X_{i,k},\, k \neq j)}
&=& \alpha\, c_i + \beta_j + \displaystyle\sum_{k=1:\, k \neq j}^p \gamma_{j,k}\, x_{i,k}.
\ee
Thus,
among all respondents $i$ with $c_i=1$,
the log odds of the conditional probability of $X_{i,j} = 1$ is decreased by $\alpha$ when $\alpha < 0$ and increased by $\alpha$ when $\alpha > 0$,
ceteris paribus.
%The effects $\alpha\, c_i$ and $\beta_j$ may be interpreted as the main effects of respondent $i$ and item $j$,
%respectively,
%whereas the effects $\gamma_{j,k}\, x_{i,k}$ may be interpreted as the interaction effects corresponding to items $j$ and $k$.

We simulate 500 data sets.
Each of the 500 data sets consists of $n = 300$ responses to $p=24$ binary items.
These 500 data sets are simulated from probability mass function \eqref{ex} with $p+\binom{p}{2}+1$ = 301 parameters $\alpha$, $\beta_j$, and $\gamma_{j,k}$.
The values of the parameters $\alpha$, $\beta_j$, and $\gamma_{j,k}$ are chosen as follows.
First,
we consider two possible values of $\alpha$:
$\alpha=0$ and $\alpha = 1/2$.
In the second scenario ($\alpha=1/2$),
we generate the covariates $c_i$ of respondents $i$ by sampling 150 out of the $n=300$ respondents at random and assigning them $c_i=0$,
while assigning all other respondents $c_i=1$.
Second,
the $p=24$ intercepts $\beta_j$ are sampled independently from Uniform$[-2,\, -1/2]$.
Last,
but not least,
the interaction weights $\gamma_{j,k}$ are generated as follows:
We sample 69 out of the $\binom{p}{2} = 276$ interaction weights $\gamma_{j,k}$ at random and assign 41 of them positive values (generated from Uniform$[1/2,\, 2]$) and 28 of them negative values (generated from Uniform$[-1,\, -1/2]$).
All other interaction weights $\gamma_{j,k}$ are set to $0$.
Given the parameters $\alpha$, $\beta_j$, and $\gamma_{j,k}$, 
we simulate 500 data sets from probability mass function \eqref{ex},
using $\alpha=0$ in Section \ref{correct} and $\alpha=1/2$ in Section \ref{incorrect}.
In each of the two scenarios ($\alpha=0$ and $\alpha=1/2$),
we use eLasso and the Bayesian approach to estimate the conditional independence graph under the assumption that $\alpha=0$.
In the first scenario ($\alpha=0$),
the model is estimated under the correct model specification,
whereas in the second scenario ($\alpha=1/2$),
the model is estimated under an incorrect model specification.
The simulation results based on these two scenarios are reviewed in Sections \ref{correct} and \ref{incorrect}.

\subsection{Simulation study with correct model specification}
\label{correct}

In the first simulation study,
we generate 500 data sets from probability mass function \eqref{ex} with $\alpha=0$ and estimate the conditional independence graph of the generalized Ising model by eLasso and the Bayesian approach under the assumption that $\alpha = 0$ (as described above).
In other words,
we estimate the conditional independence graph under the correct model specification.

First,
we assess how well the Bayesian approach can recover the conditional independence graph compared with eLasso,
using the same criteria as \citet{nature14}:
the true-positive rate (TPR) and the true-negative rate (TNR),
and the Rand index \citep{Ra71}.
The TPR, TNR, and Rand index are defined by
\be
\label{tpr}
\mbox{TPR} 
&=& \dfrac{TP}{TP + FN},
\ee
\be
\label{tnr}
\mbox{TNR} 
&=& \dfrac{TN}{TN + FP},
\ee
and
\be
\label{ri}
\mbox{Rand index} 
&=& \dfrac{TP + TN}{TP + TN + FP + FN},
\ee
respectively.
Here, 
TP denotes the number of true-positive edges; 
FP denotes the number of false-positive edges; 
TN denotes the number of true-negative edges;
and FN denotes the number of false-negative edges.
Table~\ref{simulaccuarcy} reports the mean of these criteria over the 500 simulated data sets.
\begin{table}
\begin{center}
\begin{tabular}{lcc}
  & eLasso & Bayes\\
  \hline
True-positive rate (TPR) & .227 &  .752 \\
True-negative rate (TNR) & .997 & .804\\
Rand index & .804 &  .791\\
   \hline
\end{tabular}
\caption{500 data sets were simulated from probability mass function \eqref{ex} with $\alpha=0$.
The conditional independence graph was estimated under the correct assumption that $\alpha=0$.
The performance of eLasso and the Bayesian approach is assessed in terms of the true-positive rate (TPR),
the true-negative rate (TNR),
and the Rand index,
as defined in Equations \eqref{tpr},
\eqref{tnr},
and \eqref{ri},
respectively.
These criteria are averaged over the 500 simulated data sets.}
\label{simulaccuarcy} 
\end{center}
\end{table}
Both approaches have a high Rand index,
which measures the overall accuracy of the two approaches.
That said,
there are noticable differences in the recovery of the conditional independence graph by the two approaches:
While both approaches have a high true-negative rate,
the true-positive rate of eLasso ($.227$) is substantially lower than the true-positive rate of the Bayesian approach ($.752$),
suggesting that eLasso's selection of the regularization parameter\linebreak 
\citep[based on the EBIC of][]{ChCh08} in combination with the so-called AND rule \citep[based on][]{MeBu06} may result in too sparse conditional independence graphs.
It is worth repeating that we have used here the same settings as \citet{nature14},
which are the default settings in {\tt R} package {\tt IsingFit} version 0.3.1 \citep{isingfit}.

\begin{figure}[htbp]
\begin{center}
\includegraphics[scale = 0.51]{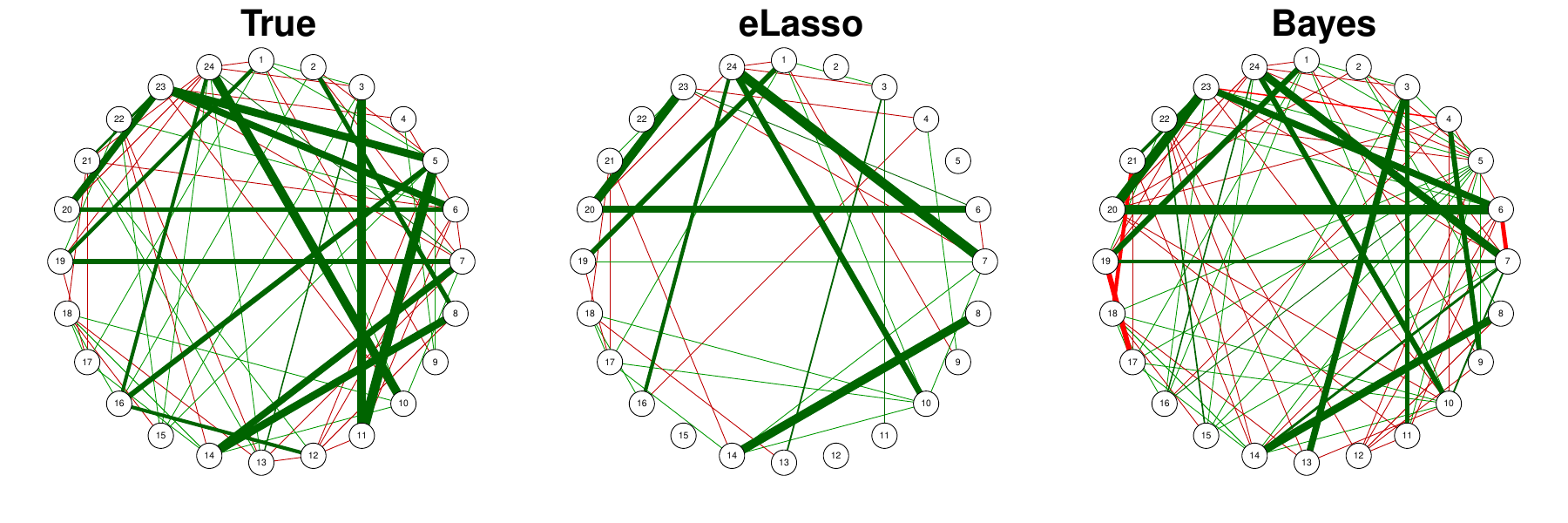}
\end{center}
\caption[]{A single example data was simulated from probability mass function \eqref{ex} with $\alpha=0$.
The conditional independence graph was estimated under the correct assumption that $\alpha=0$.
An edge between two distinct items $j$ and $k$ indicates that items $j$ and $k$ interact, 
that is,
$\gamma_{j,k} \neq 0$.
Green-colored edges represent positive interactions ($\gamma_{j,k} > 0$),
whereas red-colored edges represent negative interactions ($\gamma_{j,k} < 0$).
The width of an edge between two distinct items $j$ and $k$ is proportional to the strength of the interaction in terms of $|\gamma_{j,k}|$.
}
\label{simulnetwork}
\end{figure}

Second,
to gain more insight into how the Bayesian approach compares with eLasso in terms of graph recovery,
we simulate a single example data set from probability mass function \eqref{ex} with $\alpha=0$ and compare the true and estimated conditional independence graphs obtained by eLasso and the Bayesian approach in more detail.
Figure~\ref{simulnetwork} shows the true and estimated conditional independence graphs by eLasso and the Bayesian approach and underscores the conservative nature of eLasso (with default options):
eLasso recovers 30 of the 69 edges in the true conditional independence graph,
whereas the Bayesian approach recovers 60 of the 69 edges.

\subsection{Simulation study with incorrect model specification: omitted covariate}
\label{incorrect}

In the second simulation study,
we generate 500 data sets from probability mass function \eqref{ex} with $\alpha=1/2$ and estimate the conditional independence graph of the generalized Ising model by eLasso and the Bayesian approach under the incorrect assumption that $\alpha = 0$ (as described above).
In other words,
we estimate the conditional independence graph under an incorrect model specification.

To assess how well the Bayesian approach can recover the conditional independence graph compared with eLasso when the model is misspecified,
we report the true-positive rate (TPR), 
the true-negative rate (TNR), 
and the Rand index in Table~\ref{simulaccuarcy2},
averaged over the 500 simulated data sets.
\begin{table}[t]
\begin{center}
\begin{tabular}{lcc}
  & eLasso & Bayes\\
  \hline
True-positive rate (TPR) & .206  & .738 \\
True-negative rate (TNR) & .997 & .786 \\
Rand index & .799 & .774 \\
   \hline
\end{tabular}  
\end{center}
\caption{500 data sets were simulated from probability mass function \eqref{ex} with $\alpha=1/2$.
The conditional independence graph was estimated under the incorrect assumption that $\alpha=0$.
The performance of eLasso and the Bayesian approach is assessed in terms of the true-positive rate (TPR),
the true-negative rate (TNR),
and the Rand index,
as defined in Equations \eqref{tpr},
\eqref{tnr},
and \eqref{ri},
respectively.
These criteria are averaged over the 500 simulated data sets.}
\label{simulaccuarcy2} 
\end{table}
According to Table \ref{simulaccuarcy2}, 
the Bayesian approach is more robust against model misspecification due to omitted covariates than eLasso:
The true-positive rate of eLasso drops from .227 (correct model specification, see Table \ref{simulaccuarcy}) to .206 (incorrect model specification, see Table \ref{simulaccuarcy2}),
which is a reduction of more than 9\%.
By contrast,
the true-positive rate of the Bayesian approach drops from .752 (correct model specification, see Table \ref{simulaccuarcy}) to .738 (incorrect model specification, see Table \ref{simulaccuarcy2}),
which is a reduction of less than 2\%.
%> .206 / .227
%[1] 0.907489
%> .738 / .752
%[1] 0.981383
In other words,
not only does the Bayesian approach appear to have a substantially higher true-positive rate than eLasso,
but the true-positive rate of the Bayesian approach also appears to be less affected by model misspecification due to omitted covariates than the true-positive rate of eLasso.
%It is worth noting, however, that the false-negative rate of the Bayesian approach is more affected by model misspecification than the false-negative rate of 

\begin{figure}[htbp]
\begin{center}
\includegraphics[scale = 0.51]{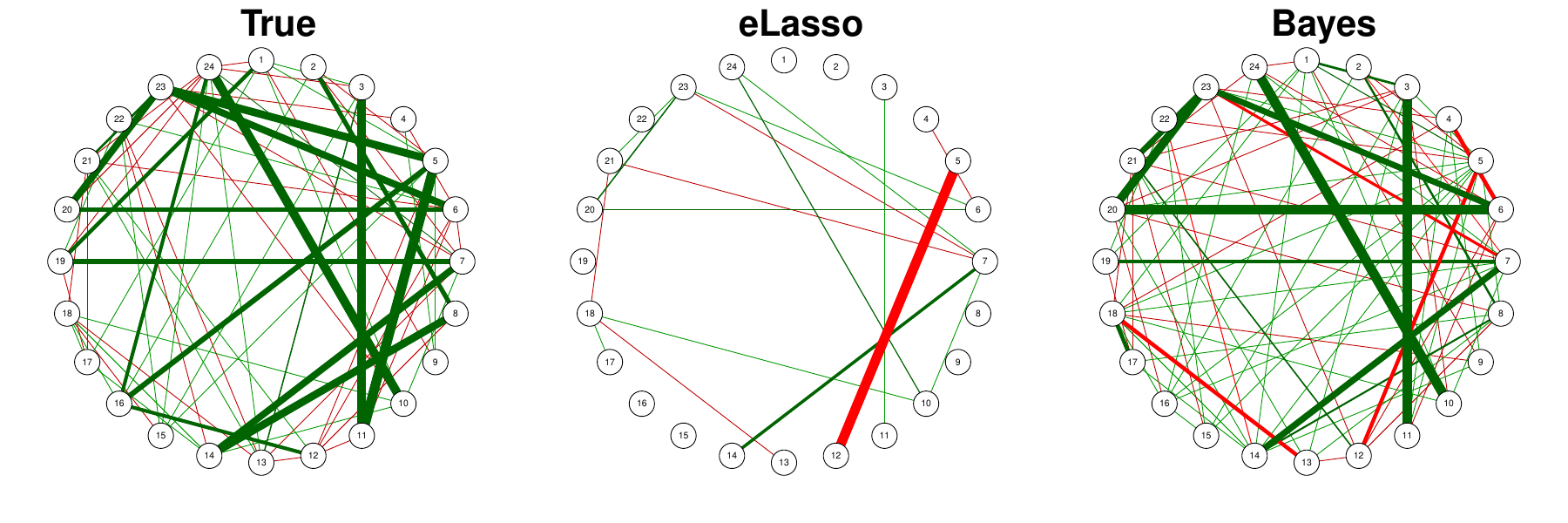}
\end{center}
\caption[]{A single example data was simulated from probability mass function \eqref{ex} with $\alpha=1/2$.
The conditional independence graph was estimated under the incorrect assumption that $\alpha=0$.
An edge between two distinct items $j$ and $k$ indicates that items $j$ and $k$ interact,
that is,
$\gamma_{j,k} \neq 0$.
Green-colored edges represent positive interactions ($\gamma_{j,k} > 0$),
whereas red-colored edges represent negative interactions ($\gamma_{j,k} < 0$).
The width of an edge between two distinct items $j$ and $k$ is proportional to the strength of the interaction in terms of $|\gamma_{j,k}|$.
}
\label{simulnetwork2}
\end{figure}

\begin{figure}[htbp]
\begin{center}
\includegraphics[scale = 0.7]{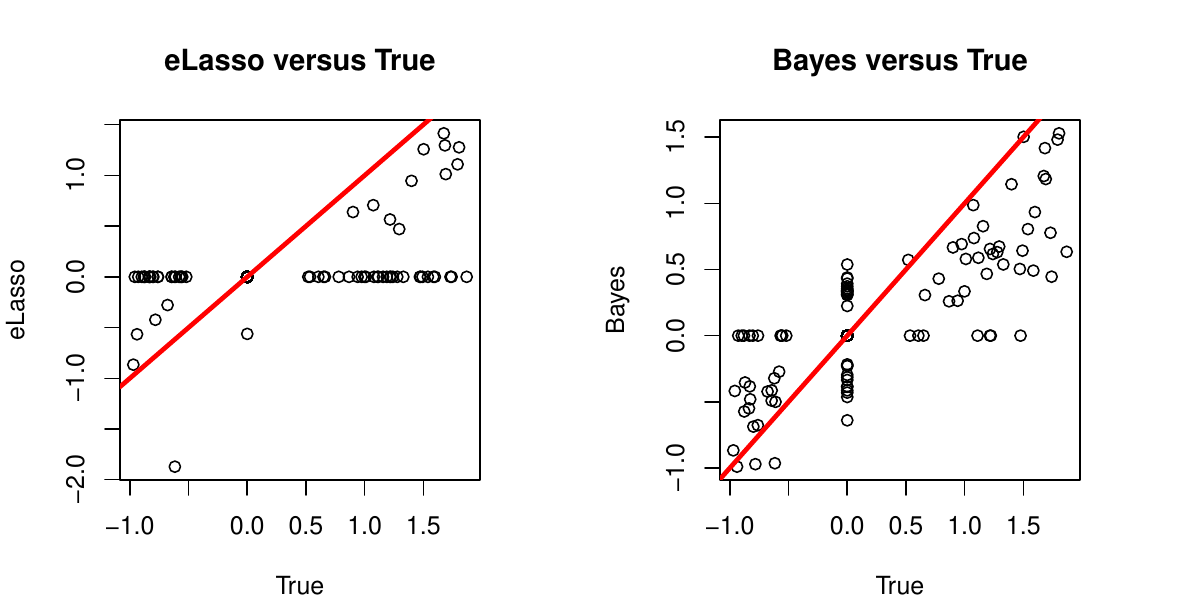}
\end{center}
\caption[]{500 data sets were simulated from probability mass function \eqref{ex} with $\alpha=1/2$.
The conditional independence graph was estimated under the incorrect assumption that $\alpha=0$.
The estimated interaction weights $\widehat\gamma_{j,k}$ are plotted against the true interaction weights $\gamma_{j,k}$.
The red-colored line is the identity line.
}
\label{simulcoeff2}
\end{figure}

Last,
but not least,
we simulate a single example data set from probability mass function \eqref{ex} with $\alpha=1/2$ to gain more insight into how the Bayesian approach compares to eLasso in terms of graph recovery.
Figure \ref{simulnetwork2} compares the true conditional independence graph to the conditional independence graphs estimated by eLasso and the Bayesian approach and highlights the advantage of the Bayesian approach over eLasso when the model is misspecified:
eLasso recovers 16 of the 69 edges in the true conditional independence graph,
whereas the Bayesian approach recovers 52 of the 69 edges.
Figure~\ref{simulcoeff2} compares the true interaction weights $\gamma_{j,k}$ and the estimated interaction weights $\widehat\gamma_{j,k}$,
with the Bayesian approach outperforming eLasso.
To quantify the advantage of the Bayesian approach over eLasso,
we compute the root mean-squared error of the estimators of the true interaction weights $\gamma_{j,k}$:
\beno
\mbox{RMSE}
&=& \sqrt{\displaystyle\sum_{j<k}^p\, (\widehat\gamma_{j,k}-{\gamma}_{j,k})^2}.
\ee
Here, 
$\widehat\gamma_{j,k}$ is the estimate of the data-generating interaction weight $\gamma_{j,k}$,
which is either the $\ell_1$-penalized logistic regression estimate (eLasso) or the posterior mean provided $\gamma_{j,k}$ is estimated to be non-zero (Bayesian approach).
It is worth noting that eLasso computes two estimates of $\gamma_{j,k}$:
one estimate based on the $\ell_1$-penalized logistic regression of item response $X_{i,j}$ on $X_{i,k}$ and all other item responses,
and one estimate based on the $\ell_1$-penalized logistic regression of item response $X_{i,k}$ on $X_{i,j}$ and all other item responses ($i = 1, \dots, n$).
These two estimates of $\gamma_{j,k}$ can differ.
To report a single estimate of $\gamma_{j,k}$,
eLasso averages over these two estimates,
and we follow eLasso here.
The RMSE of eLasso turns out to be 23.73,
whereas the RMSE of the Bayesian approach is 13.64. 
By comparison,
if the model specification is correct,
the RMSE of eLasso and the Bayesian approach are 15.58 and 18.57, 
respectively.
In other words,
when the model specification is correct, 
eLasso may have a small advantage over the Bayesian approach,
but when the model specification is incorrect,
the Bayesian approach appears to outperform eLasso in terms of RMSE.

\section{Applications to educational data}
\label{app}

We compare the Bayesian approach to eLasso by using three educational data sets.
In each application,
we compare the conditional independence graphs estimated by eLasso and the Bayesian approach,
and assess the goodness-of-fit of the estimated models.
We first provide some background on how to assess goodness-of-fit in Section \ref{sec:gof} and then present the three applications in Sections \ref{sec:data1}---\ref{sec:data3}.

\subsection{Goodness-of-fit statistics}
\label{sec:gof}

To assess the goodness-of-fit (GOF) of the estimated models,
we simulate 1,000 data sets from the estimated models.

In the case of eLasso,
we simulate model-based predictions of item responses based on the estimates of the intercepts $\beta_j$ and the interaction weights $\gamma_{j,k}$ reported by eLasso.
It is worth repeating that eLasso computes two estimates of $\gamma_{j,k}$:
one estimate based on the $\ell_1$-penalized logistic regression of item response $X_{i,j}$ on $X_{i,k}$ and all other item responses,
and one estimate based on the $\ell_1$-penalized logistic regression of item response $X_{i,k}$ on $X_{i,j}$ and all other item responses ($i = 1, \dots, n$).
To report a single estimate of $\gamma_{j,k}$,
eLasso averages over these two estimates,
and we follow eLasso here.
In the case of the Bayesian approach,
we generate posterior predictions.

We compare the simulated and observed data in terms of two statistics:
(1) $\sum_{i=1}^{n}\, x_{i,j}$,
which measures the easiness of item $j$ (intercept);
and (2) $\sum_{i=1}^{n} x_{i,j}\, x_{i,k}$,
which measures the strength of the interaction of two distinct items $j$ and $k$ (interactions).
These two statistics are sufficient statistics for the intercepts $\beta_j$ and interaction weights $\gamma_{j,k}$.
In addition,
we assess the GOF of the estimated models in terms of higher-order statistics.
To do so,
we first compute, 
for each respondent $i$,
an item-item graph where two distinct items $j$ and $k$ are connected by an edge if and only if respondent $i$ gave correct responses to both items $j$ and $k$.
The item-item graphs of respondents should not be confused with the conditional independence graph of the Ising model:
Item-item graphs represent data structure (the responses of respondents to pairs of items),
whereas the conditional independence graph of the Ising model represents model structure (the conditional independence structure of the Ising model).
We then compute the number of cliques of size $l$ in the item-item graph of each of the $n$ respondents,
and average the number of cliques of size $l$ over the $n$ respondents.
A clique of size $l$ is a maximal complete subset of nodes,
that is, 
a subset of $l$ nodes such that all $\binom{l}{2}$ pairs of nodes are connected by edges and it is impossible to add nodes without losing the property of completeness \citep{La96}.
These goodness-of-fit statistics are used in the three applications in Sections \ref{sec:data1}---\ref{sec:data3}.

\subsection{Abortion data}
\label{sec:data1}

\begin{figure}[htbp]
\begin{center}
\includegraphics[scale = 0.55]{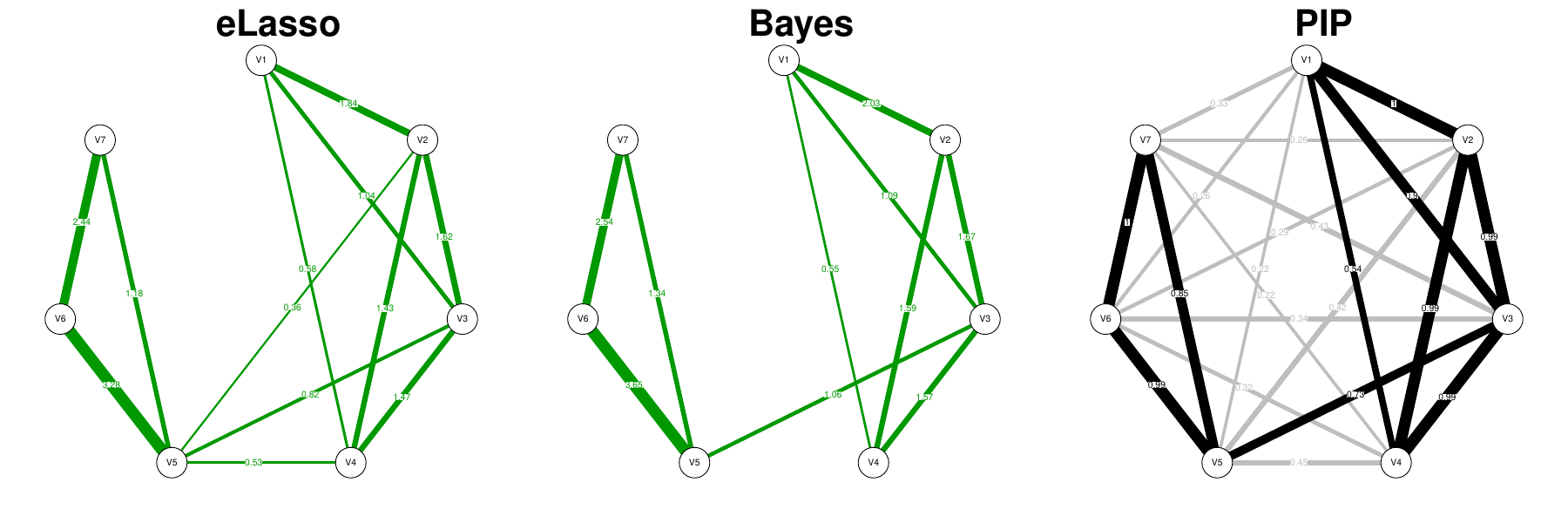}
\end{center}
\caption[]{Abortion data: Conditional independence graphs estimated by eLasso and the Bayesian approach based on a Markov chain Monte Carlo sample of size 10,000.
An edge between two distinct items $j$ and $k$ indicates that items $j$ and $k$ interact, 
that is,
$\gamma_{j,k} \neq 0$.
Green-colored edges represent positive interactions ($\gamma_{j,k} > 0$),
whereas red-colored edges represent negative interactions ($\gamma_{j,k} < 0$).
The width of an edge between two distinct items $j$ and $k$ is proportional to the strength of the interaction in terms of $|\gamma_{j,k}|$.
The graph labeled PIP shows the posterior interaction probabilities of pairs of distinct items $j$ and $k$,
that is,
the posterior probability of the event that the indicator $\lambda_i$ corresponding to the interaction weight $\gamma_{j,k}$ equals $1$.
}
\label{Abortionnetwork}
\end{figure}

We first use a classic data set consisting of data on attitudes towards abortion,
collected by the British Social Attitudes Survey Panel between 1983 and 1986 \citep{social1987}. 
Respondents were asked whether abortion should be allowed by law under the following circumstances:
\begin{enumerate}
\item The woman decides on her own whether she does not wish to have the child. 
\item The couple agrees that they do not wish to have the child.
\item The woman is not married and does not wish to marry the man.
\item The couple cannot afford any more children.
\item There is a strong chance of a defect in the baby.
\item The woman’s health is seriously endangered by the pregnancy.
\item The woman became pregnant as a result of rape.
\end{enumerate}
The data correspond to binary responses (either 1=``yes'' or 0=``no'') by $n = 642$ individuals to the $p = 7$ items described above. 
The resulting Ising model has $p + \binom{p}{2} = 28$ parameters.

When applied to the abortion data,
eLasso takes about 3/10 seconds,
whereas the Bayesian approach takes about 5.4 minutes.
Implementation details are provided in Sections \ref{db} and \ref{sec:impl}.
\begin{figure}[t]
\begin{center}
\includegraphics[scale = 0.7]{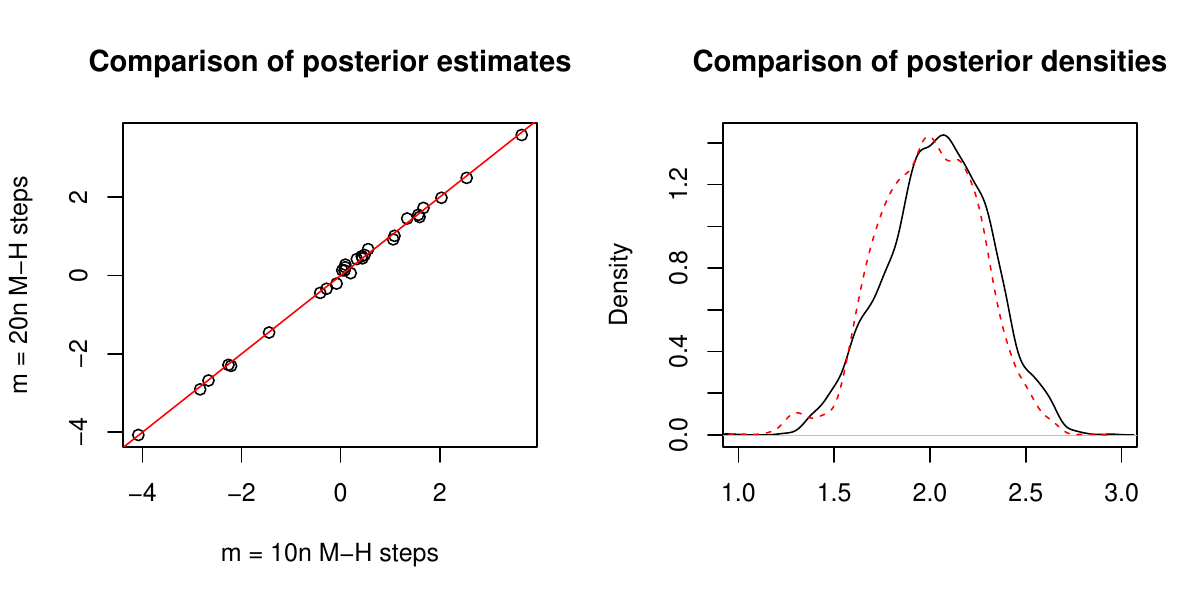}
\end{center}
\caption[]{Abortion data.
Left: posterior means of all $\theta_i$'s based on $m = 10\, n$ and $m = 20\, n$ Metropolis-Hastings (M-H) steps in Step 2 of Algorithm~\ref{spikedouble Metropolis-Hastings algorithmalg}.
Right: posterior density of $\gamma_{1,2}$.
The solid and dotted lines indicate posterior densities obtained based on $m = 10\, n$ and $m = 20\, n$ Metropolis-Hastings (M-H) steps,
respectively.}
\label{Abortioninner}
\end{figure}
We first assess the performance of the Bayesian algorithm as a function of $m$ by running it with $m = 10\, n$ and $m = 20\, n$ in Step 2 of Algorithm~\ref{spikedouble Metropolis-Hastings algorithmalg}.
Figure~\ref{Abortioninner} indicates that posterior means and posterior distributions do not change much as $m$ is increased.

\begin{figure}[htbp]
\begin{center}
\includegraphics[scale = 0.8]{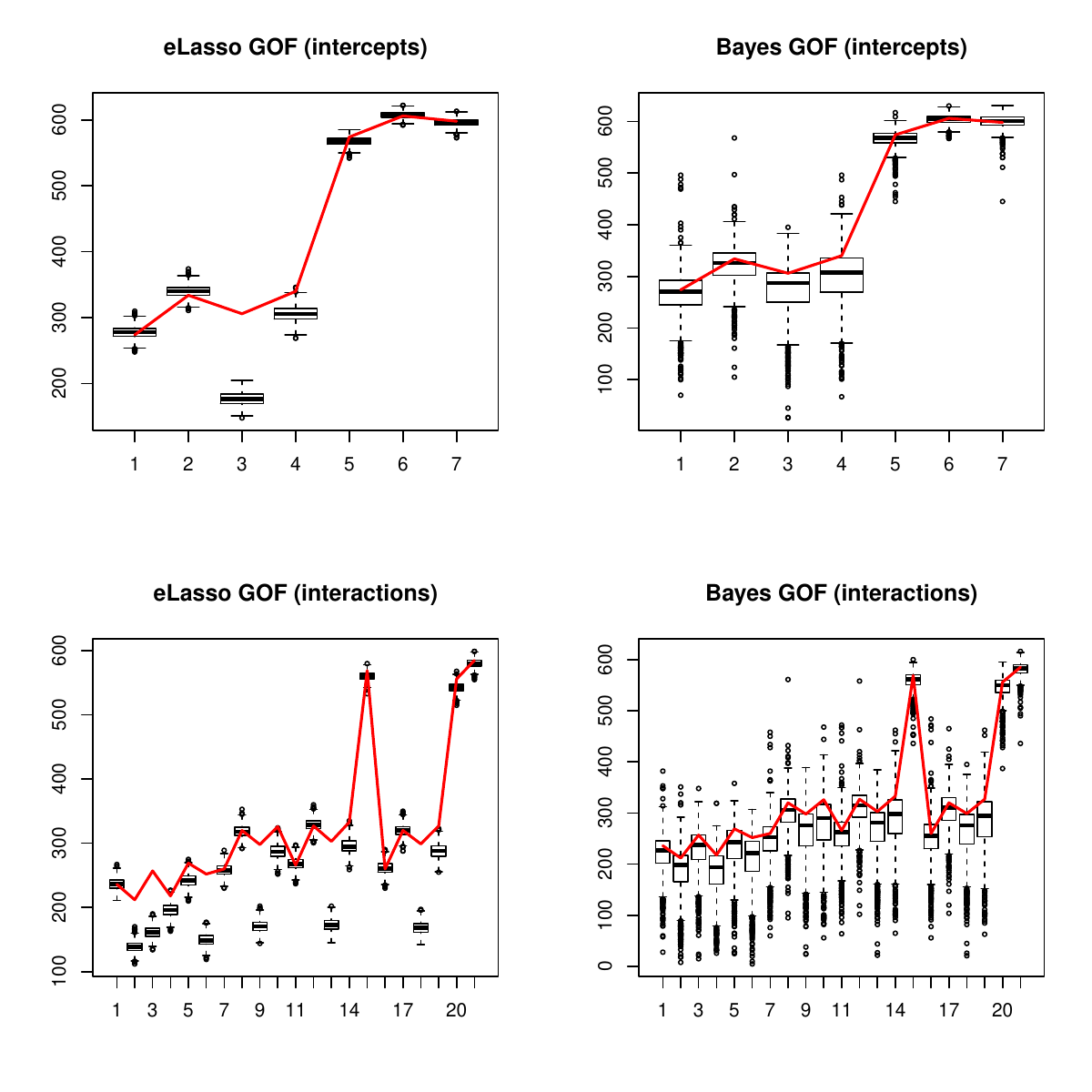}
\end{center}
\caption[]{Abortion data: 
GOF assessment in terms of sufficient statistics for the intercepts $\beta_j$ and the interaction weights $\gamma_{j,k}$ based on 1,000 simulated data sets.
The sufficient statistics are stated in Section \ref{sec:gof}.
The red lines indicate the observed values of the sufficient statistics.}
\label{Abortiongof}
\end{figure}

\begin{figure}[htbp]
\begin{center}
\includegraphics[scale = .45]{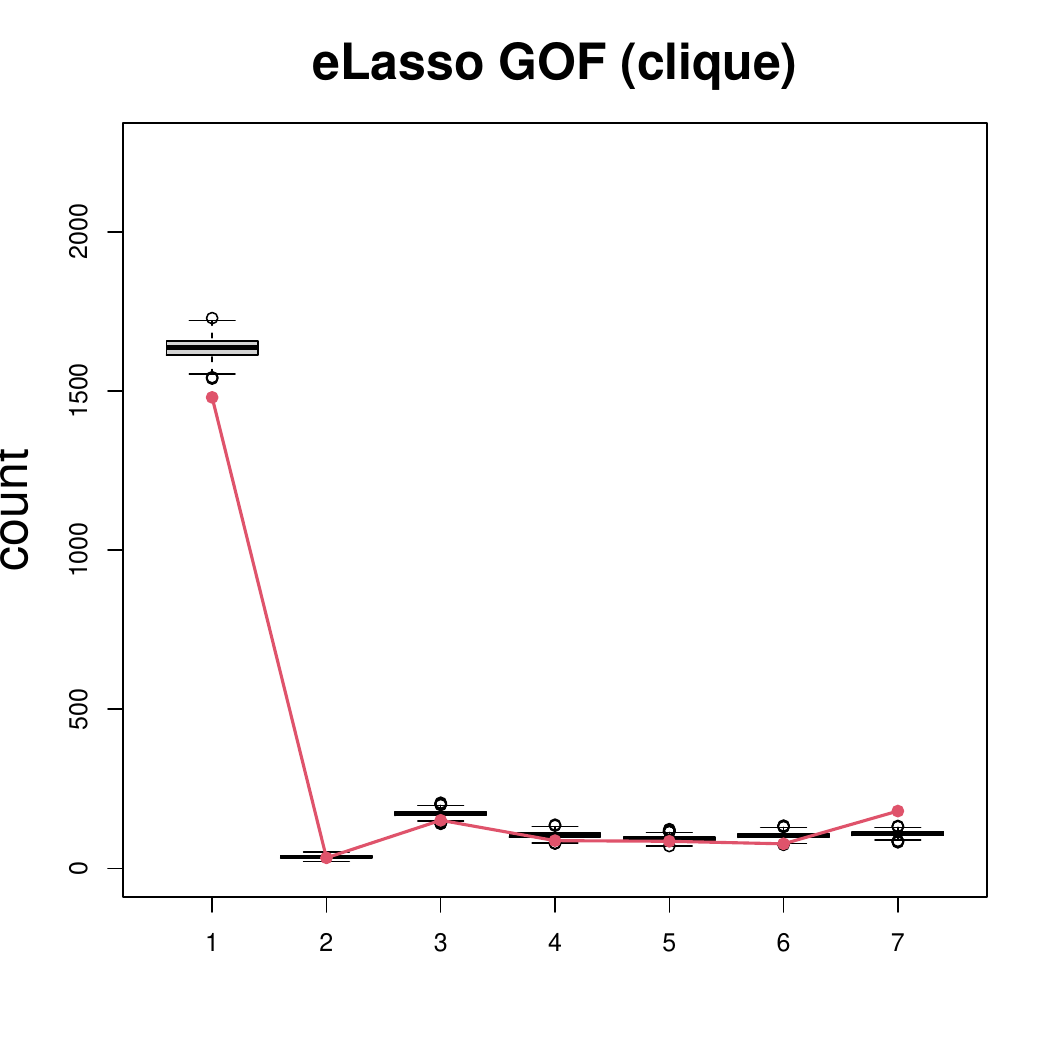}
\includegraphics[scale = .45]{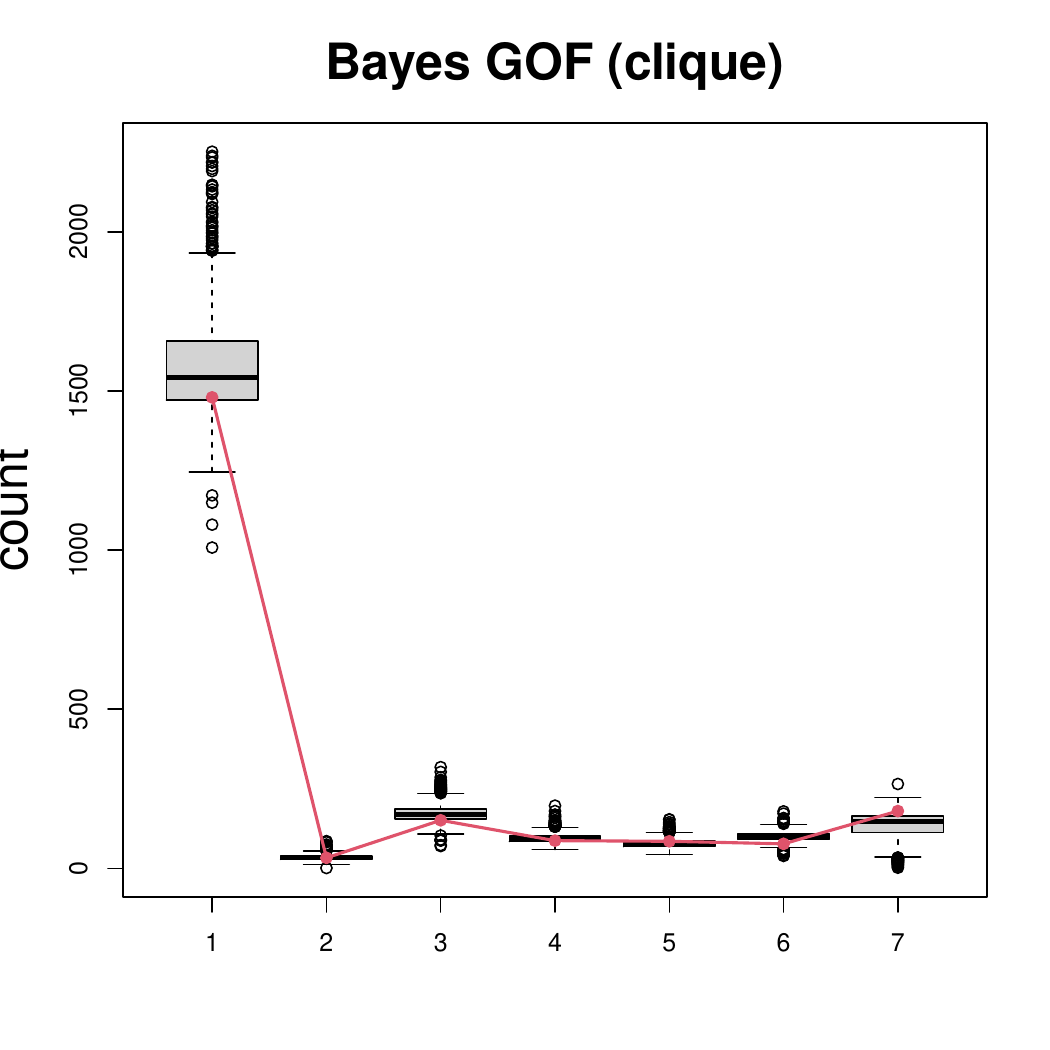}
\end{center}
\caption[]{Abortion data: 
GOF assessment in terms of cliques based on 1,000 simulated data sets.
For each simulated data set and each respondent,
the number of cliques of size $l$ in the simulated item-item graph is computed,
and then summed over all $n$ respondents.
The red lines indicate the number of cliques of size $l$ in the observed item-item graphs,
summed over the $n$ respondents.}
\label{Abortiongof2}
\end{figure}
The conditional independence graphs estimated by eLasso and the Bayesian approach are represented in Figure~\ref{Abortionnetwork}.
To compare them,
note that \citet{jeon2020} detected two groups of items in the abortion data set by using latent space item response models: 
The first group of items (items 1--4) measures whether women may abort for personal reasons,
whereas the second group of items (items 5--7) measures whether women may abort for medical or other reasons.
According to Figure \ref{Abortionnetwork},
both eLasso and the Bayesian approach agree with the observation of \citet{jeon2020} that there are two groups of items with more connections within groups than between groups,
but the Bayesian approach shows a more clear-cut separation of the two groups of items compared with eLasso.
To shed light onto the difference between eLasso and the Bayesian approach,
we inspect the posterior interaction probabilities of pairs of distinct items $j$ and $k$,
that is,
the posterior probability of the event that the indicator $\lambda_i$ corresponding to the interaction weight $\gamma_{j,k}$ equals $1$.
Comparing the eLasso estimate of the conditional independence graph to the posterior interaction probabilities reveals that eLasso seems to connect all pairs of items with a posterior interaction probability of at least $.42$.
By contrast,
the Bayesian approach connects pairs of items with a posterior interaction probability of at least 1/2.
In other words,
it seems that eLasso is liberal compared with the Bayesian approach,
at least if the two-group structure found by \citet{jeon2020} is accepted as a reference point.
It is worth pointing out that the true conditional independence graph is unknown,
but the two-group structure found by \citet{jeon2020} makes sense,
because the first group of items is concerned with abortion for personal reasons,
whereas the second group of items is concerned with abortion for medical and other reasons.

Last,
but not least,
we assess the GOF of the estimated models.
Figure~\ref{Abortiongof} suggests that the Bayesian approach outperforms eLasso in terms of sufficient statistics for the interaction weights $\gamma_{j,k}$.
Figure \ref{Abortiongof2} reveals that both the Bayesian approach and eLasso match the observed number of cliques rather well,
although the Bayesian approach may have a small advantage over eLasso with respect to cliques of sizes 1 and 7.
Here,
as in Sections \ref{sec:data2} and \ref{sec:data3},
there appears to be more variation in the model-based predictions of the Bayesian approach compared with eLasso.
The reason is that the Bayesian approach takes into account the uncertainty about the parameters and averages over all parameters,
whereas eLasso does not.

\subsection{Deductive Reasoning Verbal (DRV) data}
\label{sec:data2}

The Competence Profile Test of Deductive Reasoning Verbal (DRV: \citealp{spiel:01, spiel:08}) was developed based on Piaget's cognitive developmental theory \citep{piaget:71},
with a view to evaluating the cognitive development stages of children and adolescents.
The data set consists of $n = 418$ respondents and $p = 24$ items.
Some information about the items of the DRV data set can be found in the supplement and more information can be found in \cite{spiel:01} and \citet{jin2019doubly}. 
The resulting Ising model has $p + \binom{p}{2} = 300$ parameters.

\begin{figure}[t]
\begin{center}
\includegraphics[scale = 0.55]{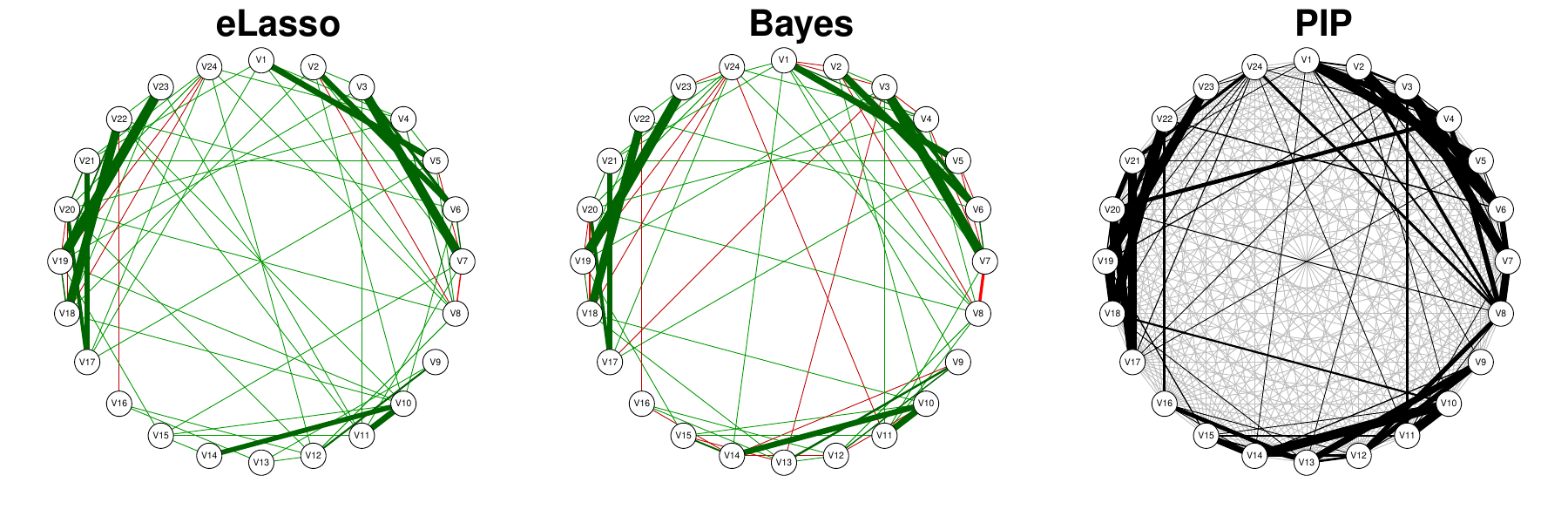}
\end{center}
\caption[]{DRV data: estimated conditional independence graphs based on eLasso and the Bayesian approach with a Markov chain Monte Carlo sample of size 20,000.
An edge between two distinct items $j$ and $k$ indicates that items $j$ and $k$ interact, 
that is,
$\gamma_{j,k} \neq 0$.
Green-colored edges represent positive interactions ($\gamma_{j,k} > 0$),
whereas red-colored edges represent negative interactions ($\gamma_{j,k} < 0$).
The width of an edge between two distinct items $j$ and $k$ is proportional to the strength of the interaction in terms of $|\gamma_{j,k}|$.
The graph labeled PIP shows the posterior interaction probabilities of pairs of distinct items $j$ and $k$,
that is,
the posterior probability of the event that the indicator $\lambda_i$ corresponding to the interaction weight $\gamma_{j,k}$ equals $1$.
}
\label{DRVnetwork}
\end{figure}

\begin{figure}[htbp]
\begin{center}
\includegraphics[scale = 0.7]{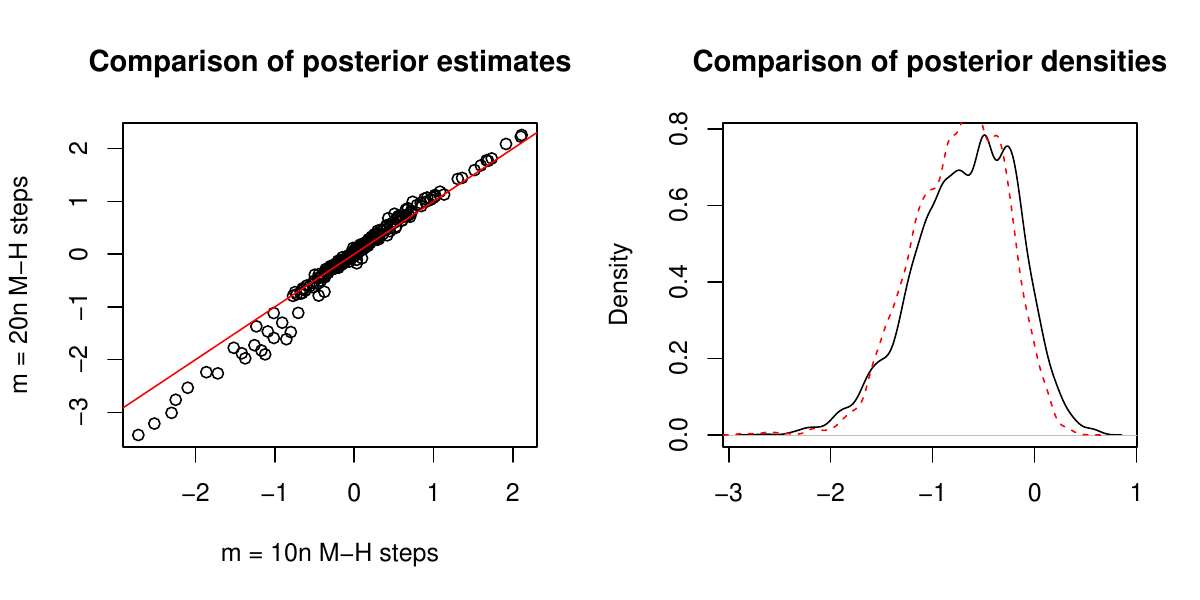}
\end{center}
\caption[]{DRV data:
Left: posterior means of all $\theta_i$'s based on $m = 10\, n$ and $m = 20\, n$ Metropolis-Hastings (M-H) steps in Step 2 of Algorithm~\ref{spikedouble Metropolis-Hastings algorithmalg}.
Right: posterior density of $\gamma_{1,2}$.
The solid and dotted lines indicate posterior densities obtained based on $m = 10\, n$ and $m = 20\, n$ Metropolis-Hastings (M-H) steps,
respectively.}
\label{DRVinner}
\end{figure}

Applied to the DRV data set,
eLasso approach takes about .94 seconds,
whereas the Bayesian approach takes about 3.5 hours.
To assess whether the number $m$ of Metropolis-Hastings steps in Step 2 of the Bayesian algorithm was large enough,
we increase $m=10\, n$ to $m=20\, n$.
Figure~\ref{DRVinner} suggests that the results do not change too much when $m$ is increased from $m=10\, n$ to $m=20\, n$. 
The estimated conditional independence graphs obtained by eLasso and the Bayesian approach are shown in Figure~\ref{DRVnetwork}. 
eLasso and the Bayesian approach agree on the signs of 244 of the 276 interaction weights $\gamma_{j,k}$.
The eLasso approach reports 209 edges,
whereas the Bayesian approach reports 203 edges.
The Bayesian approach reports more edges with negative interaction weights (22) than eLasso (8).
Although the true conditional independence graph is unknown,
it is known that some the items in the DRV test have a negative logical relationship by construction:
e.g., 
items 7 and 8 have a negative logical relationship by construction,
and both eLasso and the Bayesian approach report an edge with a negative interaction weight $\gamma_{7,8}$,
but eLasso reports fewer other negative relationships than the Bayesian approach.
 
\begin{figure}[htbp]
\begin{center}
\includegraphics[scale = 0.8]{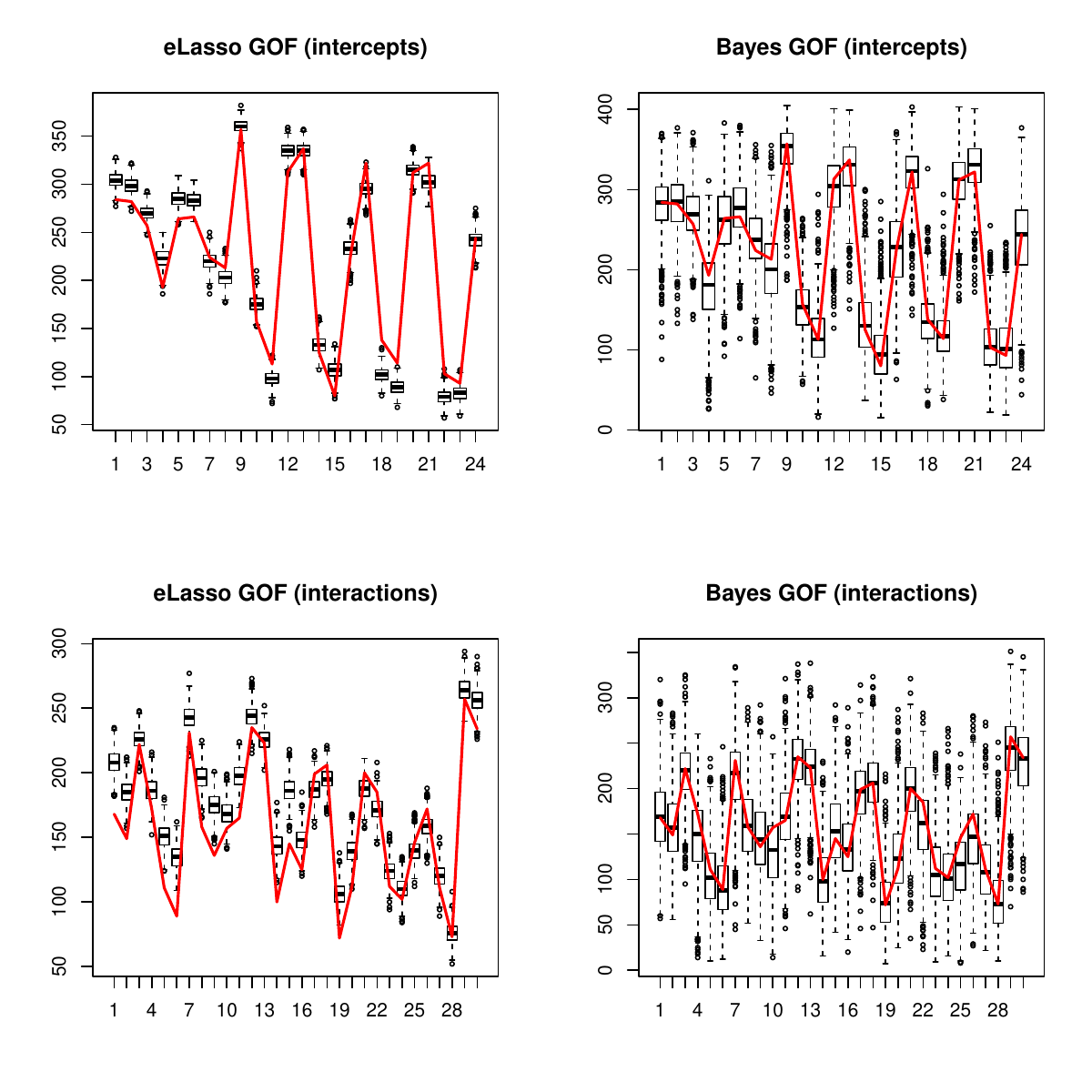}
\end{center}
\caption[]{DRV data: 
GOF assessment in terms of sufficient statistics for the intercepts $\beta_j$ and the interaction weights $\gamma_{j,k}$ based on 1,000 simulated data sets.
The sufficient statistics are stated in Section \ref{sec:gof}.
The red lines indicate the observed values of the sufficient statistics.}
\label{DRVgof}
\end{figure}

\begin{figure}[htbp]
\begin{center}
\includegraphics[scale = .45]{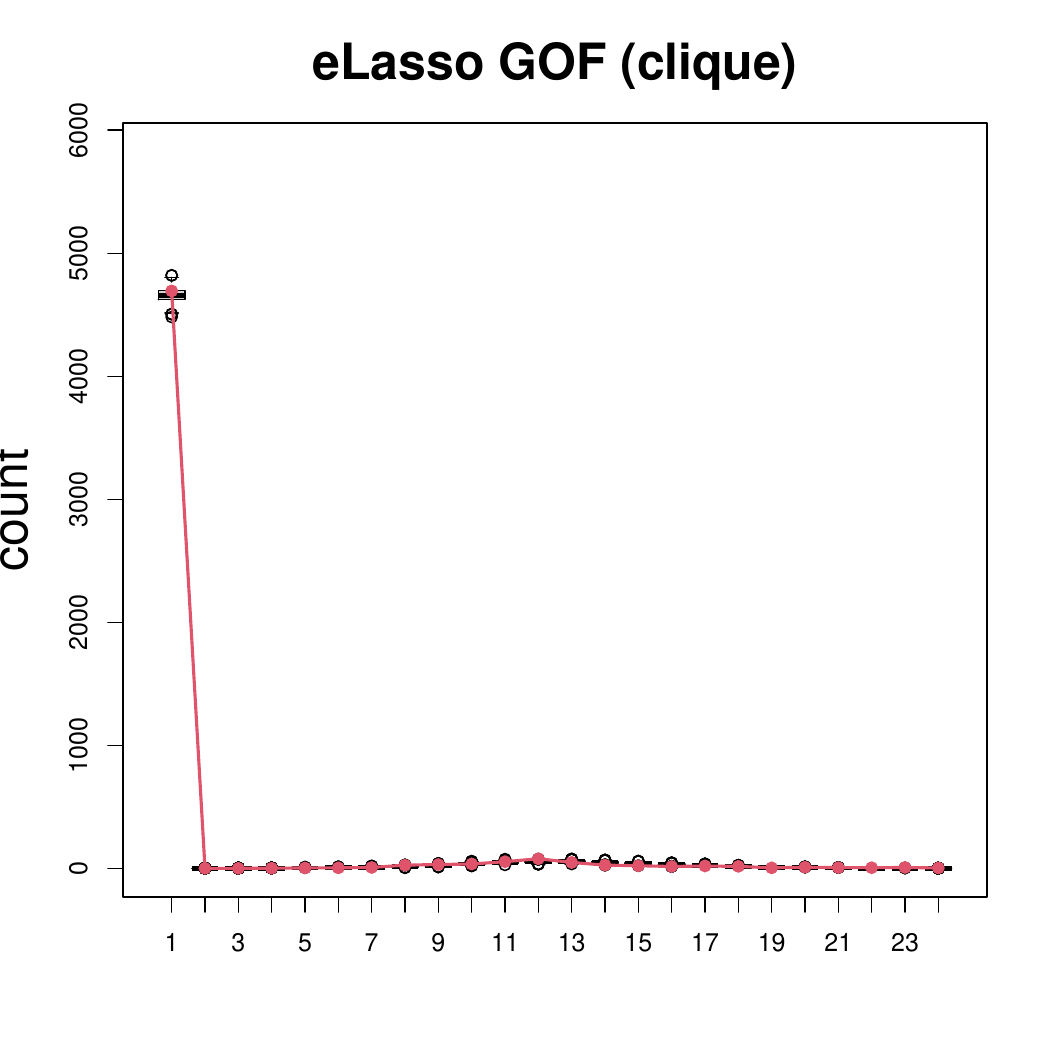}
\includegraphics[scale = .45]{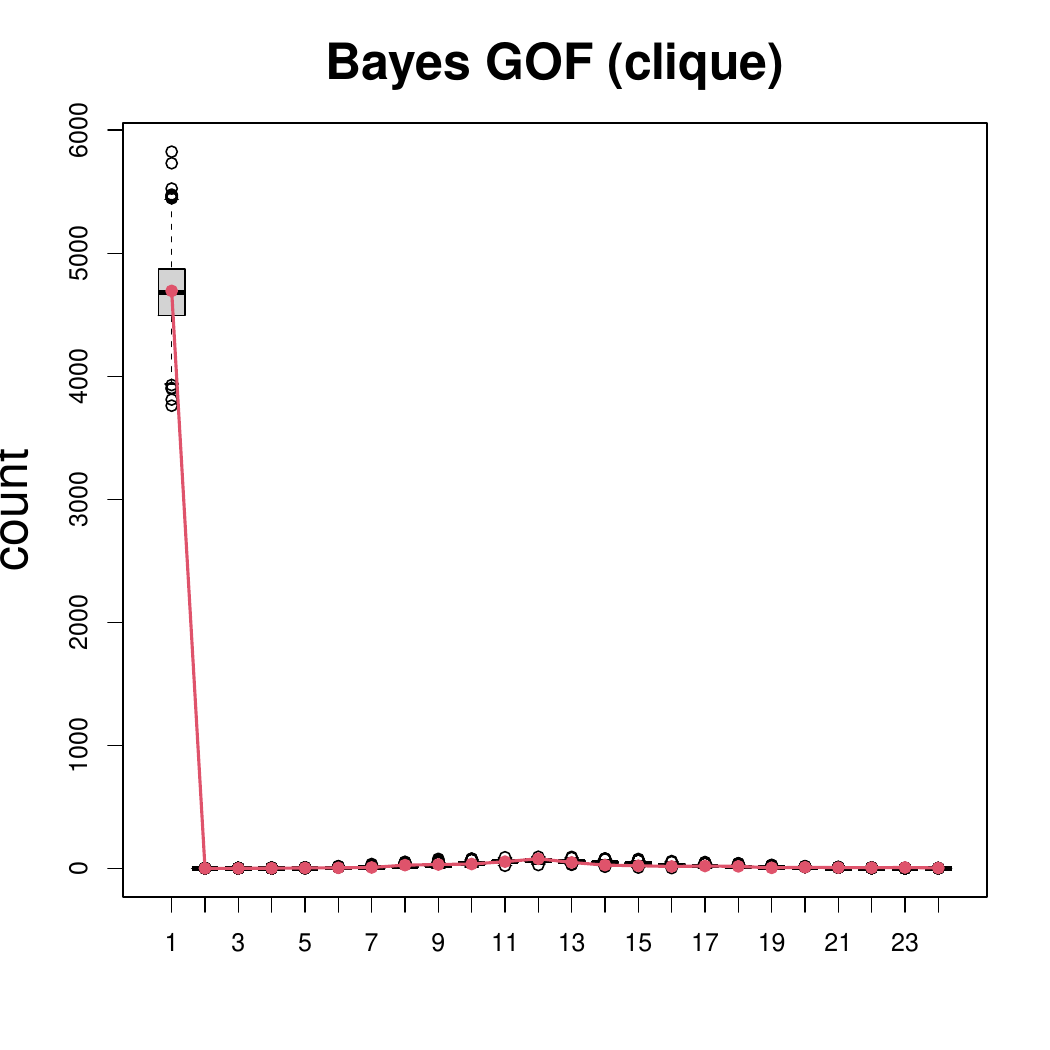}
\end{center}
\caption[]{DRV data:
GOF assessment in terms of cliques based on 1,000 simulated data sets.
For each simulated data set and each respondent,
the number of cliques of size $l$ in the simulated item-item graph is computed,
and then summed over all $n$ respondents.
The red lines indicate the number of cliques of size $l$ in the observed item-item graphs,
summed over the $n$ respondents.}
\label{drvgof2}
\end{figure}

As in the previous example, 
we assess the GOF of the estimated model.
Since we have 300 sufficient statistics for 24 intercepts and 276 interactions, 
we show sufficient statistics for the intercept and first 30 interaction weights. 
Figures \ref{DRVgof} and \ref{drvgof2} suggest that the Bayesian approach compares favorably to eLasso in terms of GOF with respect to sufficient statistics for the intercepts $\beta_j$ and the interaction weights $\gamma_{j,k}$,
and cliques.

\subsection{Korean middle school data}
\label{sec:data3}

For decades, 
the Korean public K-12 system has been criticized for creating a competitive environment that may have a negative impact on the intellectual, mental, and behavioral development of students. 
To understand the developmental status of students and evaluate the effect of the competitive environment, 
the Office of Education in Gyeongi Province (the Seoul Metropolitan area) conducted surveys to assess the mental and physical health, creativity, ethics, autonomy, and democratic conciseness of students \citep{Gyeonggi:2012}. 
The data was collected in 2014 and concern 9th grade students (3rd grade students in the Korean middle school).
All responses were transformed into binary responses:
1 (``strongly disagree''), 
2 (``disagree''), 
and 3 (``do not disagree or agree'') were transformed to 0;
and 4 (``agree'') and 5 (``strongly agree'') were transformed to 1.
We analyze the middle school data set consisting of $n=$ 3,784 respondents and $p = 70$ items. 
A list of all items can be found in the supplement.
The resulting Ising model has $p + \binom{p}{2} = $ 2,485 parameters. 

\begin{figure}[htbp]
\begin{center}
\includegraphics[ scale = 0.85]{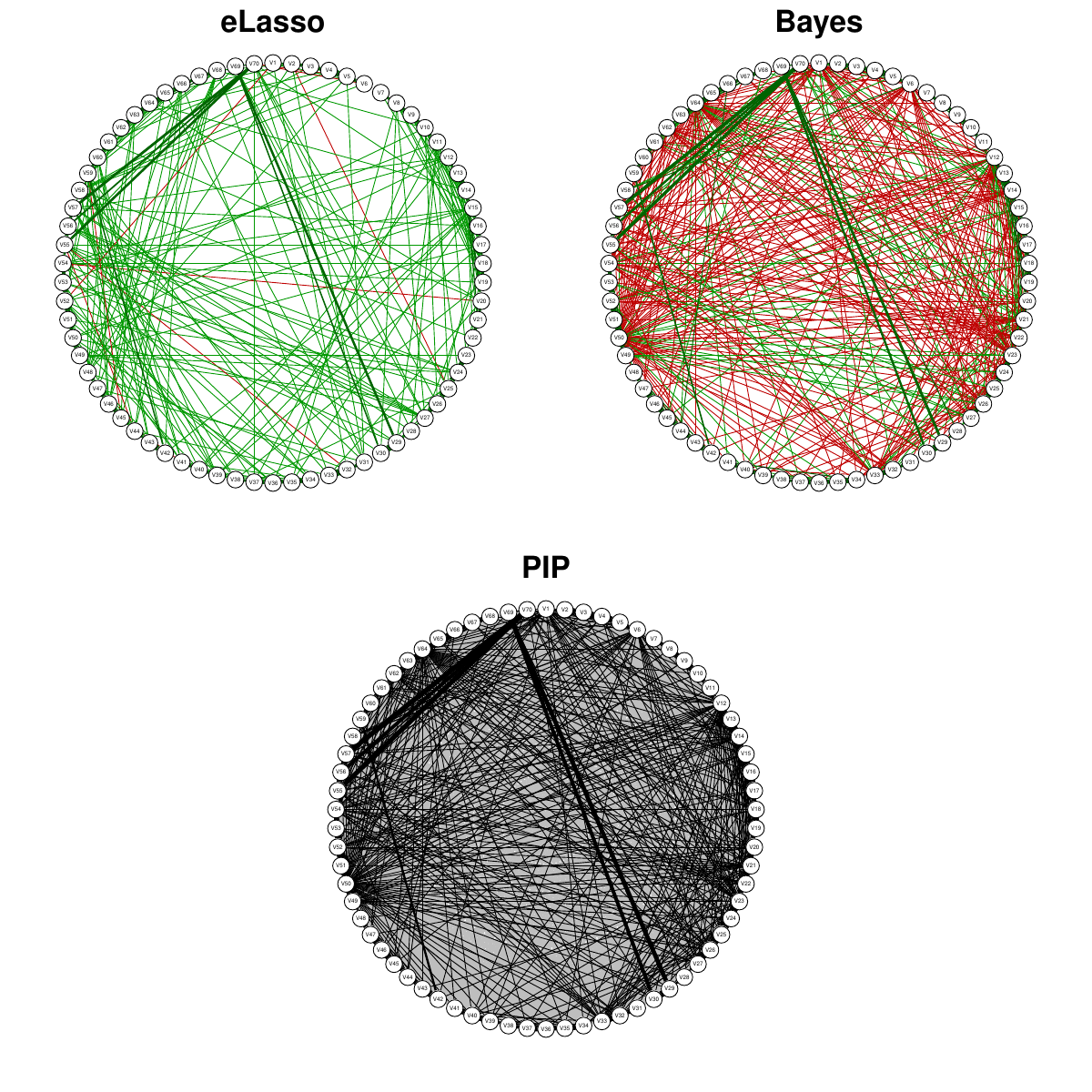}
\end{center}
\caption[]{Korean middle school data:
Conditional independence graphs estimated by eLasso and the Bayesian approach,
based on a Markov chain Monte Carlo sample of size 20,000.
An edge between two distinct items $j$ and $k$ indicates that items $j$ and $k$ interact, 
that is,
$\gamma_{j,k} \neq 0$.
Green-colored edges represent positive interactions ($\gamma_{j,k} > 0$),
whereas red-colored edges represent negative interactions ($\gamma_{j,k} < 0$).
The width of an edge between two distinct items $j$ and $k$ is proportional to the strength of the interaction in terms of $|\gamma_{j,k}|$.
The graph labeled PIP shows the posterior interaction probabilities of pairs of distinct items $j$ and $k$,
that is,
the posterior probability of the event that the indicator $\lambda_i$ corresponding to the interaction weight $\gamma_{j,k}$ equals $1$.
}
\label{Middlenetwork}
\end{figure}

The Bayesian approach takes about 430 hours,
whereas the eLasso approach takes about 24.33 seconds.
The eLasso approach sets 2,101 of the 2,415 interaction weights $\gamma_{j,k}$ to 0.
By contrast, 
the Bayesian approach sets 1,992 of the interaction weights to 0.
The eLasso approach and the Bayesian approach agree on the sign of 1,980 of the 2,415 interaction weights $\gamma_{j,k}$.
The estimated conditional independence graphs are shown in Figure~\ref{Middlenetwork}. 

\begin{table}[t]
\centering
\begin{tabular}{cccccc}
Positive  & Item name & Estimate\\
  \hline
$\gamma_{29,30}$ & ``appearance satisfaction" (29), ``appearance esteem" (30) & 3.90 \\
& & (2.16, 5.58)\\
$\gamma_{4,5}$ & ``feel lonely" (4), ``feel sad/depressed" (5) & 3.68 \\
& & (1.81, 5.69)\\
$\gamma_{42,43}$ & ``nervous when take the exam" (42), ``nervous before the exam" (43) & 3.45 \\
& & (1.76, 5.05)\\
$\gamma_{23,26}$ & ``like to hang out with others" (23), ``happy to be with someone else" (26) & 3.20 \\
& & (1.10, 5.28)\\
$\gamma_{32,33}$ & ``ability does not change" (32), ``even if I try, ability does not change" (33) & 2.98 \\
& & (.50, 5.27)\\
   \hline   
Negative  & Item name & Estimate\\
  \hline
$\gamma_{6,70}$ & ``mental ill-being" (6), ``self-esteem" (70) & -.55\\
& & (-2.72, 1.13)\\
$\gamma_{49,62}$ & ``relationship with friends" (49), ``academic stress" (62) & -.50 \\
& & (-2.65, 1.15)\\
$\gamma_{12,64}$ & ``sense of citizenship" (12), ``academic stress" (64) & -.50 \\
& & (-2.57, 1.31)\\
$\gamma_{12,61}$ & ``sense of citizenship" (12), ``academic stress" (61) & -.486 \\
& & (-2.61, 1.25)\\
$\gamma_{37,50}$ & ``self-driven learning" (37), ``relationship with friends" (50) & -.46
\\
& & (-2.46, 1.50)\\
   \hline   
\end{tabular}
\caption{Korean middle school data: the five strongest positive and negative interactions in terms of the posterior mean of the interaction weights $\gamma_{j,k}$.
The estimates mentioned above are posterior means.
The intervals are 95\% posterior credible intervals.}
%The 95\% highest posterior density is calculated by using {\tt R} package {\tt coda}.}
\label{middletable} 
\end{table}

We provide some descriptive explanations based on the top 5 strongest positive and negative interactions in terms of the posterior mean of the parameters $\gamma_{j,k}$ shown in Table \ref{middletable}. 
To interpret the results,
we present unofficial translations of the Korean items within quotation marks.
The strongest positive interaction occurs between items 29 (``I have a favorable face") and 30 ("My appearance is attractive"),
which may be unsurprising.
The second strongest positive interaction occurs between items 4 (``Sometimes I experience loneliness for no reason") and 5 (``At times I am sad and depressed for no reason"),
suggesting that loneliness is related to sadness. 
The third strongest positive interaction occurs between items 42 (``I am nervous when I try to take the exam") and 43 (``I am more nervous before the exam"). 
Both items measure test anxiety. 

The strongest negative interaction occurs between items 6 (``I sometimes want to die for no reason") and 70 (``I have a positive attitude towards myself"),
which makes sense,
because the two questions measure opposite attitudes.
The second strongest negative interaction occurs between items 49 (``I feel comfortable when I am with my school friends") and 62 (``I feel uneasy when I play with my friends"),
which likewise measure opposite attitudes.
The third strongest negative interaction occurs between items 12 (``Foreigners living in Korea should be treated in the same way as Koreans") and 64 (``I can ignore friendship to get better grades in grade or entrance exams"),
suggesting a negative association between compassion for foreigners and selfish pursuit of academic interests.

\begin{figure}[htbp]
\begin{center}
\includegraphics[scale = 0.8]{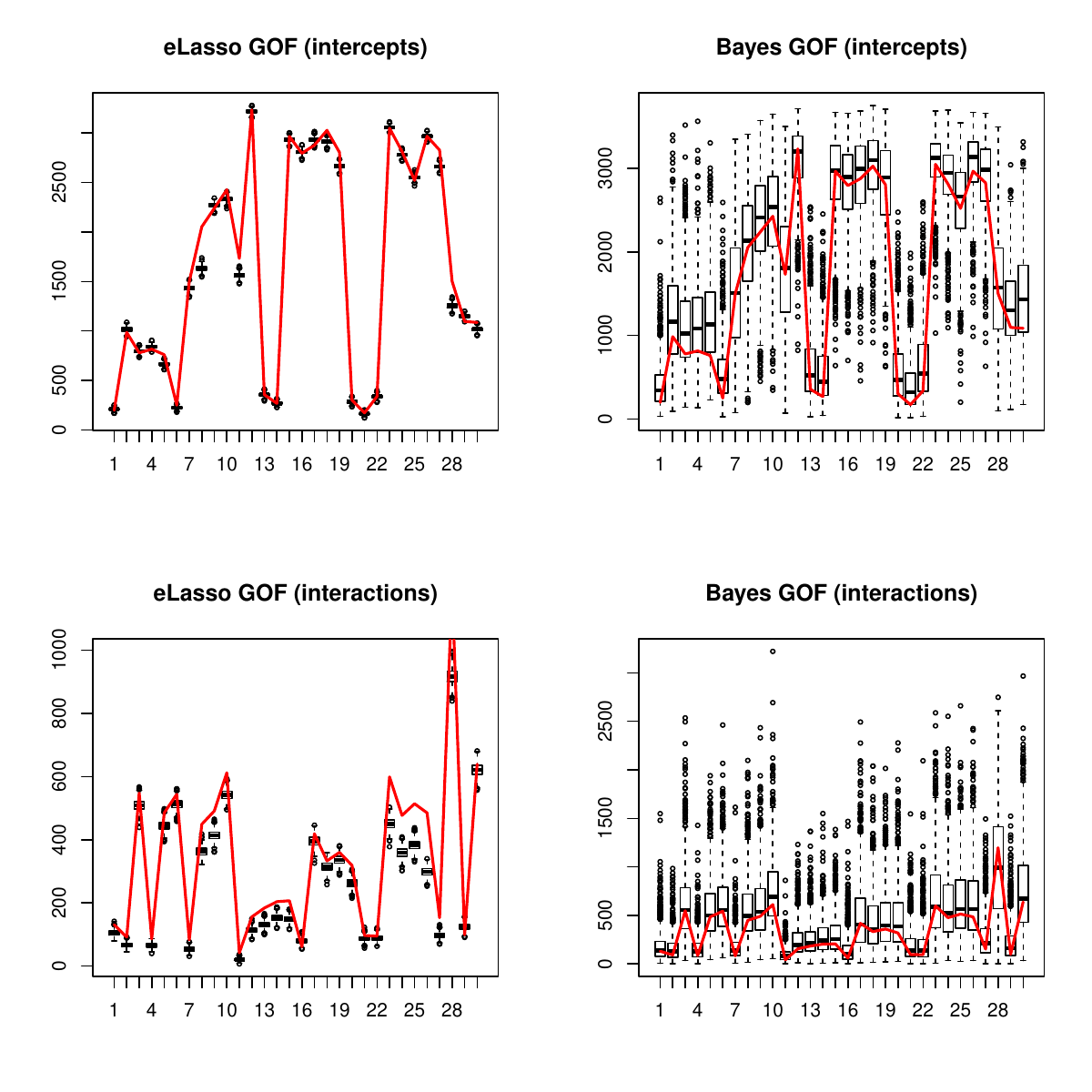}
\end{center}
\caption[]{Korean middle school data.
GOF assessment in terms of sufficient statistics for the intercepts $\beta_j$ and the interaction weights $\gamma_{j,k}$ based on 1,000 simulated data sets.
The red lines indicate the observed values of the sufficient statistics.}
\label{schoolgof}
\end{figure}

\begin{figure}[htbp]
\begin{center}
\includegraphics[scale = 0.45]{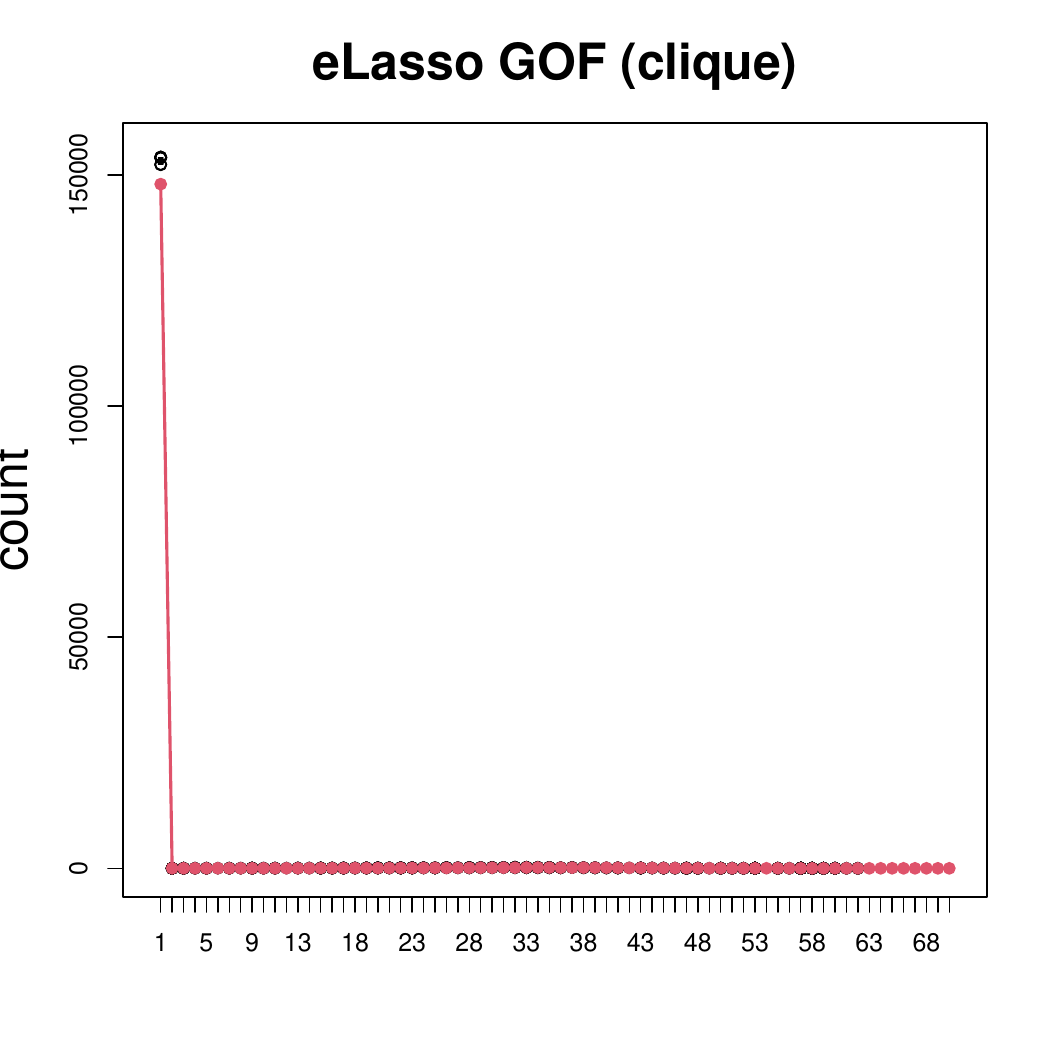}
\includegraphics[scale = 0.45]{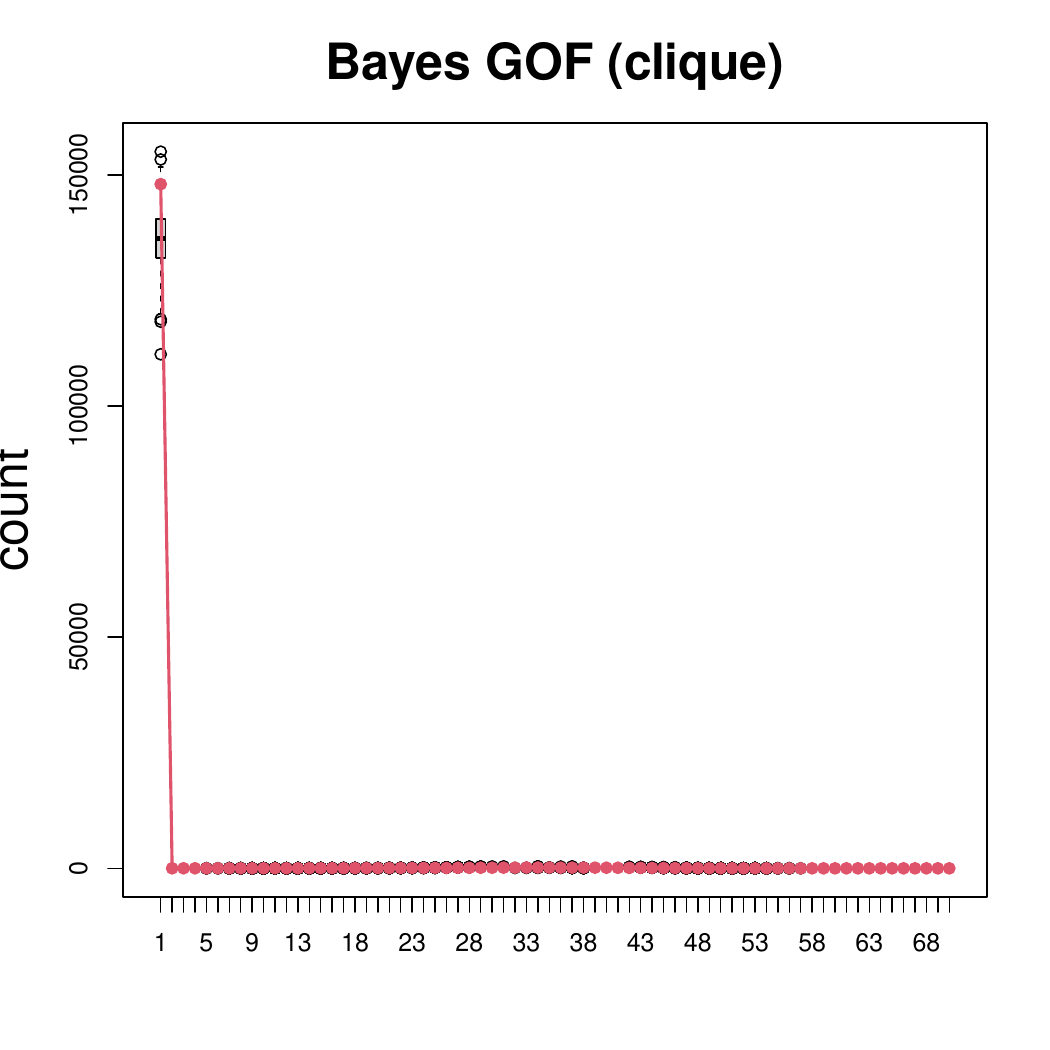}
\end{center}
\caption[]{Korean middle school data: 
GOF assessment in terms of cliques,
averaged over the $n$ item-item repondent networks of the $n$ respondents.
The red lines indicate the observed numbers of cliques,
summed over the $n$ item-item repondent networks of the $n$ respondents.}
\label{koreangof3}
\end{figure}

Last,
but not least,
we assess the GOF of the estimated models.
The figure shows the GOF of the estimated models in terms of the sufficient statistics for the first 30 intercepts and interaction weights---note that there are 2,480 parameters and hence 2,480 sufficient statistics,
and displaying the GOF of the estimated models in terms of all 2,480 sufficient statistics is too space-consuming.
Figures~\ref{schoolgof} and \ref{koreangof3} indicate that the Bayesian approach performs well in terms of GOF compared with eLasso.
%We also check whether the observed statistics belong to the interquartile range (between the first and third quartiles) of simulated statistics. 
%Of the 2,485 observed statistics, 
%2,395 statistics are included in the interquartile range by the Bayesian approach, 
%whereas only 273 statistics are included by eLasso. 
%This demonstrates that the Bayesian approach fits the data well compared with eLasso,
As before,
there appears to be more variation in the model-based predictions of the Bayesian approach compared with eLasso,
arising from the fact that the posterior predictions take into account the uncertainty about the parameters,
whereas the model-based predictions of eLasso do not.

\section{Discussion}
\label{disc}

We have developed a Bayesian approach for Ising models with doubly-intractable posterior distributions,
with applications to educational data.
The proposed approach helps quantify the uncertainty about the estimated conditional independence graph along with the parameters of the model,
and appears to be more robust against model misspecification due to omitted covariates than the $\ell_1$-penalized nodewise logistic regression approach.

To address the statistical and computational challenges arising from doubly-intractable posterior distributions,
we have combined two approaches: (1) a double Metropolis-Hastings algorithm algorithm \citep{liang2010double} and (2) stochastic search variable selection methods \citep{George:93,Ishwaran:05}. 
We note that the proposed Bayesian approach is inexact,
in the sense that the stationary distribution of the Markov chains constructed by the proposed Bayesian algorithm is not the desired target posterior. 
The reason is that the double Metropolis-Hastings algorithm algorithm generates an auxiliary variable from an approximate distribution in Step 2 of Algorithm~\ref{spikedouble Metropolis-Hastings algorithmalg} on page \pageref{algorithm1}. 
However,
the double Metropolis-Hastings algorithm is feasible in high-dimensional settings with thousands of parameters,
whereas many alternatives are not.
Several approaches have been developed to reduce variance,
but are inexact.
For example, 
\cite{alquier2014noisy} and \citet{stoehr2017noisy} provide Hamiltonian variants of double Metropolis-Hastings algorithm, and \cite{friel2016exploiting} develops control variates for intractable likelihood functions. 
Developing an exact algorithm for such models is still an open question (leaving aside perfect sampling, which can be expensive in terms of computing time) 

It is worth noting that there variations on the approach proposed here,
depending on the choice of the variable selection method \citep[see, e.g.,][]{o2009review}. 
For instance, 
instead of using the vector of indicators $\boldsymbol{\lambda}$ in the model, 
Bayesian lasso methods \citep{park2008bayesian,yi2008bayesian} directly approximate the spike and slab shape of the prior on the model parameters $\boldsymbol{\theta}$. 
The horseshoe prior \citep{carvalho2010horseshoe} is a promising alternative.

An open issue is the scalability of the Bayesian algorithm,
that is,
the ability of the Bayesian algorithm to scale up to larger data sets with more respondents $n$ or more items $p$.
While we were able to apply the Bayesian algorithm to Ising models with $p + \binom{p}{2} = $ 2,485 parameters based on item responses from $n=$ 3,784 respondents (Section \ref{sec:data3}),
the computing time required to obtain samples from an approximation to the posterior distribution (430 hours, that is, almost 18 days) suggests that more work is needed to scale up the Bayesian algorithm to larger $n$ and larger $p$.
One of the main computational bottlenecks is the generation of auxiliary variables.
There are a number of ideas for addressing these computational challenges. 
For instance, 
\cite{park2019function} propose a function emulation approach that replaces expensive importance sampling schemes with fast Gaussian process approximations. 
\cite{bouranis2017efficient} provides a practical Bayesian approach for large networks by correcting Markov chain Monte Carlo samples from pseudo-posterior distribution. 
These and other ideas---and combinations of them---constitute interesting avenues for future research.

\section*{Supplementary materials}

The supplement provides more background on the data sets used in Section \ref{app}.
All data and all source code used in the paper can be downloaded from https://github.com/jwpark88/itemBayes.

\section*{Acknowledgements}

Jaewoo Park was partially supported by the Yonsei University Research Fund of 2019-22-0194 and the National Research Foundation of Korea (NRF-2020R1C1C1A0100386811). Ick Hoon Jin was partially supported by the Yonsei University Research Fund of 2019-22-0210 and the National Research Foundation of Korea (NRF-2020R1A2C1A01009881). 
Michael Schweinberger was partially supported by the U.S.\ National Science Foundation (NSF award DMS-1812119). 
The authors are grateful to an anonymous associate editor and two anonymous reviewers,
whose constructive comments have greatly improved the paper.

\bibliography{base,Reference}

\begin{thebibliography}{}

\bibitem[\protect\citeauthoryear{{Agresti}}{{Agresti}}{2002}]{Ag02}
{Agresti}, A. (2002).
\newblock {\em Categorical Data Analysis\/} (2 ed.).
\newblock Hoboken: John Wiley \& Sons.

\bibitem[\protect\citeauthoryear{Alquier, Friel, Everitt, and Boland}{Alquier
  et~al.}{2016}]{alquier2014noisy}
Alquier, P., N.~Friel, R.~Everitt, and A.~Boland (2016).
\newblock Noisy {M}onte {C}arlo: Convergence of {M}arkov chains with
  approximate transition kernels.
\newblock {\em Statistics and Computing\/}~{\em 26\/}(1-2), 29--47.

\bibitem[\protect\citeauthoryear{Anandkumar, Tan, Huang, and
  Willsky}{Anandkumar et~al.}{2012}]{Anetal12}
Anandkumar, A., V.~Y.~F. Tan, F.~Huang, and A.~S. Willsky (2012).
\newblock High-dimensional structure estimation in {I}sing models: {L}ocal
  separation criterion.
\newblock {\em The Annals of Statistics\/}~{\em 40}, 1346--1375.

\bibitem[\protect\citeauthoryear{Atchad{\'e}}{Atchad{\'e}}{2006}]{atchade2006adaptive}
Atchad{\'e}, Y.~F. (2006).
\newblock An adaptive version for the {M}etropolis adjusted {L}angevin
  algorithm with a truncated drift.
\newblock {\em Methodology and Computing in Applied Probability\/}~{\em
  8\/}(2), 235--254.

\bibitem[\protect\citeauthoryear{Atchade, Lartillot, and Robert}{Atchade
  et~al.}{2013}]{AtLaRo12}
Atchade, Y.~F., N.~Lartillot, and C.~Robert (2013).
\newblock Bayesian computation for statistical models with intractable
  normalizing constants.
\newblock {\em Brazilian Journal of Probability and Statistics\/}~{\em
  27\/}(4), 416--436.

\bibitem[\protect\citeauthoryear{Beaumont, Zhang, and Balding}{Beaumont
  et~al.}{2002}]{beaumont2002approximate}
Beaumont, M.~A., W.~Zhang, and D.~J. Balding (2002).
\newblock Approximate {B}ayesian computation in population genetics.
\newblock {\em Genetics\/}~{\em 162\/}(4), 2025--2035.

\bibitem[\protect\citeauthoryear{Besag}{Besag}{1974}]{besag1974spatial}
Besag, J. (1974).
\newblock Spatial interaction and the statistical analysis of lattice systems.
\newblock {\em Journal of the Royal Statistical Society. Series B
  (Methodological)\/}~{\em 36}, 192--236.

\bibitem[\protect\citeauthoryear{{Besag}}{{Besag}}{1974}]{Bj74}
{Besag}, J. (1974).
\newblock Spatial interaction and the statistical analysis of lattice systems.
\newblock {\em Journal of the Royal Statistical Society, Series B\/}~{\em 36},
  192--225.

\bibitem[\protect\citeauthoryear{Borsboom}{Borsboom}{2008}]{Bo08}
Borsboom, D. (2008).
\newblock Psychometric perspectives on diagnostic systems.
\newblock {\em Journal of Clinical Psychology\/}~{\em 64\/}(9), 1089--1108.

\bibitem[\protect\citeauthoryear{Bouranis, Friel, and Maire}{Bouranis
  et~al.}{2017}]{bouranis2017efficient}
Bouranis, L., N.~Friel, and F.~Maire (2017).
\newblock Efficient {B}ayesian inference for exponential random graph models by
  correcting the pseudo-posterior distribution.
\newblock {\em Social Networks\/}~{\em 50}, 98--108.

\bibitem[\protect\citeauthoryear{Bouranis, Friel, and Maire}{Bouranis
  et~al.}{2018}]{bouranis2018bayesian}
Bouranis, L., N.~Friel, and F.~Maire (2018).
\newblock Bayesian model selection for exponential random graph models via
  adjusted pseudolikelihoods.
\newblock {\em Journal of Computational and Graphical Statistics\/}~{\em
  27\/}(3), 516--528.

\bibitem[\protect\citeauthoryear{Bresler and Karzand}{Bresler and
  Karzand}{2020}]{BrKa20}
Bresler, G. and M.~Karzand (2020).
\newblock Learning a tree-structured {I}sing model in order to make
  predictions.
\newblock {\em The Annals of Statistics\/}~{\em 48}, 713--737.

\bibitem[\protect\citeauthoryear{B\"uhlmann and van~de Geer}{B\"uhlmann and
  van~de Geer}{2011}]{BuGe11}
B\"uhlmann, P. and S.~van~de Geer (2011).
\newblock {\em Statistics for High-Dimensional Data: Methods, Theory and
  Applications}.
\newblock New York: Springer-Verlag.

\bibitem[\protect\citeauthoryear{Butts}{Butts}{2018}]{Bu12}
Butts, C.~T. (2018).
\newblock A perfect sampling method for exponential family random graph models.
\newblock {\em The Journal of Mathematical Sociology\/}~{\em 42\/}(1), 17--36.

\bibitem[\protect\citeauthoryear{Caimo and Friel}{Caimo and
  Friel}{2011}]{caimo2011bayesian}
Caimo, A. and N.~Friel (2011).
\newblock Bayesian inference for exponential random graph models.
\newblock {\em Social Networks\/}~{\em 33\/}(1), 41--55.

\bibitem[\protect\citeauthoryear{Caimo and Friel}{Caimo and
  Friel}{2013}]{caimo2013bayesian}
Caimo, A. and N.~Friel (2013).
\newblock Bayesian model selection for exponential random graph models.
\newblock {\em Social Networks\/}~{\em 35\/}(1), 11--24.

\bibitem[\protect\citeauthoryear{Caimo and Friel}{Caimo and
  Friel}{2014}]{bergm.jss}
Caimo, A. and N.~Friel (2014).
\newblock {{Bergm}: {B}ayesian Exponential Random Graphs in R}.
\newblock {\em Journal of Statistical Software\/}~{\em 61}, 1--25.

\bibitem[\protect\citeauthoryear{Caimo and Gollini}{Caimo and
  Gollini}{2020}]{CaGo20m}
Caimo, A. and I.~Gollini (2020).
\newblock A multilayer exponential random graph modelling approach for weighted
  networks.
\newblock {\em Computational Statistics \& Data Analysis\/}~{\em 142},
  106--125.

\bibitem[\protect\citeauthoryear{Carvalho, Polson, and Scott}{Carvalho
  et~al.}{2010}]{carvalho2010horseshoe}
Carvalho, C.~M., N.~G. Polson, and J.~G. Scott (2010).
\newblock The horseshoe estimator for sparse signals.
\newblock {\em Biometrika\/}~{\em 97\/}(2), 465--480.

\bibitem[\protect\citeauthoryear{Chatterjee}{Chatterjee}{2007}]{Ch07}
Chatterjee, S. (2007).
\newblock Stein's method for concentration inequalities.
\newblock {\em Probability Theory and Related Fields\/}~{\em 138}, 305--321.

\bibitem[\protect\citeauthoryear{Chen and Chen}{Chen and Chen}{2008}]{ChCh08}
Chen, J. and Z.~Chen (2008).
\newblock Extended {B}ayesian information criteria for model selection with
  large model spaces.
\newblock {\em Biometrika\/}~{\em 95}, 759--771.

\bibitem[\protect\citeauthoryear{Eddelbuettel, Fran{\c{c}}ois, Allaire,
  Chambers, Bates, and Ushey}{Eddelbuettel et~al.}{2011}]{eddelbuettel2011rcpp}
Eddelbuettel, D., R.~Fran{\c{c}}ois, J.~Allaire, J.~Chambers, D.~Bates, and
  K.~Ushey (2011).
\newblock Rcpp: Seamless {R} and {C}++ integration.
\newblock {\em Journal of Statistical Software\/}~{\em 40\/}(8), 1--18.

\bibitem[\protect\citeauthoryear{Epskamp, Borsboom, and Fried}{Epskamp
  et~al.}{2018}]{Epskamp:2018}
Epskamp, S., D.~Borsboom, and E.~I. Fried (2018).
\newblock Estimating psychological networks and their accuracy: A tutorial
  paper.
\newblock {\em Behavior Research Methods\/}~{\em 50}, 195--212.

\bibitem[\protect\citeauthoryear{Everitt}{Everitt}{2012}]{Ev12}
Everitt, R.~G. (2012).
\newblock Bayesian parameter estimation for latent {M}arkov random fields and
  social networks.
\newblock {\em Journal of Computational and Graphical Statistics\/}~{\em 21},
  940--960.

\bibitem[\protect\citeauthoryear{Flegal, Haran, and Jones}{Flegal
  et~al.}{2008}]{flegal2008markov}
Flegal, J.~M., M.~Haran, and G.~L. Jones (2008).
\newblock {M}arkov chain {M}onte carlo: Can we trust the third significant
  figure?
\newblock {\em Statistical Science\/}~{\em 23}, 250--260.

\bibitem[\protect\citeauthoryear{Frank and {Strauss}}{Frank and
  {Strauss}}{1986}]{FoSd86}
Frank, O. and D.~{Strauss} (1986).
\newblock Markov graphs.
\newblock {\em Journal of the American Statistical Association\/}~{\em 81},
  832--842.

\bibitem[\protect\citeauthoryear{Friel, Mira, Oates, et~al.}{Friel
  et~al.}{2016}]{friel2016exploiting}
Friel, N., A.~Mira, C.~J. Oates, et~al. (2016).
\newblock Exploiting multi-core architectures for reduced-variance estimation
  with intractable likelihoods.
\newblock {\em Bayesian Analysis\/}~{\em 11\/}(1), 215--245.

\bibitem[\protect\citeauthoryear{George and McCulloch}{George and
  McCulloch}{1993}]{George:93}
George, E. and R.~McCulloch (1993).
\newblock Variable selection via {G}ibbs sampling.
\newblock {\em Journal of the American Statistical Association\/}~{\em 88},
  881--889.

\bibitem[\protect\citeauthoryear{Ghosal and Mukherjee}{Ghosal and
  Mukherjee}{2020}]{GhMu20}
Ghosal, P. and S.~Mukherjee (2020).
\newblock Joint estimation of parameters in {I}sing model.
\newblock {\em The Annals of Statistics\/}~{\em 48}, 785--810.

\bibitem[\protect\citeauthoryear{Goldstein}{Goldstein}{2015}]{goldstein2015compartmental}
Goldstein, J. (2015).
\newblock {\em Compartmental, spatial and point process models for infectious
  diseases}.
\newblock Ph.\ D. thesis, the Pennsylvania State University.

\bibitem[\protect\citeauthoryear{{Gyeonggi Provincial Office of
  Education}}{{Gyeonggi Provincial Office of Education}}{2012}]{Gyeonggi:2012}
{Gyeonggi Provincial Office of Education} (2012).
\newblock {\em Plan of innovation school management}.
\newblock Republic of Korea: Gyeonggi Province.

\bibitem[\protect\citeauthoryear{Hunter and Handcock}{Hunter and
  Handcock}{2012}]{hunter2012inference}
Hunter, D.~R. and M.~S. Handcock (2012).
\newblock Inference in curved exponential family models for networks.
\newblock {\em Journal of Computational and Graphical Statistics\/}.

\bibitem[\protect\citeauthoryear{Hunter, Krivitsky, and Schweinberger}{Hunter
  et~al.}{2012}]{HuKrSc12}
Hunter, D.~R., P.~N. Krivitsky, and M.~Schweinberger (2012).
\newblock Computational statistical methods for social network models.
\newblock {\em Journal of Computational and Graphical Statistics\/}~{\em 21},
  856--882.

\bibitem[\protect\citeauthoryear{Ishwaran and Rao}{Ishwaran and
  Rao}{2005}]{Ishwaran:05}
Ishwaran, H. and J.~S. Rao (2005).
\newblock Spike and slab variable selection: {F}requentist and {B}ayesian
  strategies.
\newblock {\em The Annals of Statistics\/}~{\em 33}, 730--773.

\bibitem[\protect\citeauthoryear{Ising}{Ising}{1925}]{Ising}
Ising, E. (1925).
\newblock Beitrag zur {T}heorie des {F}erromagnetismus.
\newblock {\em Zeitschrift f\"ur Physik {A}\/}~{\em 31}, 253--258.

\bibitem[\protect\citeauthoryear{Jeon, Jin, Schweinberger, and Baugh}{Jeon
  et~al.}{2021}]{jeon2020}
Jeon, M., I.~H. Jin, M.~Schweinberger, and S.~Baugh (2021).
\newblock Mapping unobserved item-respondent interactions: A latent space item
  response model with interaction map.
\newblock {\em Psychometrika\/}.
\newblock To appear.

\bibitem[\protect\citeauthoryear{Jin and Jeon}{Jin and
  Jeon}{2019}]{jin2019doubly}
Jin, I.~H. and M.~Jeon (2019).
\newblock A doubly latent space joint model for local item and person
  dependence in the analysis of item response data.
\newblock {\em Psychometrika\/}~{\em 84\/}(1), 236--260.

\bibitem[\protect\citeauthoryear{Jin, Yuan, and Liang}{Jin
  et~al.}{2013}]{JiYuLi13}
Jin, I.~H., Y.~Yuan, and F.~Liang (2013).
\newblock Bayesian analysis for exponential random graph models using the
  adaptive exchange sampler.
\newblock {\em Statistics and its Interface\/}~{\em 6}, 559--576.

\bibitem[\protect\citeauthoryear{Jones, Haran, Caffo, and Neath}{Jones
  et~al.}{2006}]{jones2006fixed}
Jones, G.~L., M.~Haran, B.~S. Caffo, and R.~Neath (2006).
\newblock Fixed-width output analysis for {M}arkov chain {M}onte {C}arlo.
\newblock {\em Journal of the American Statistical Association\/}~{\em
  101\/}(476), 1537--1547.

\bibitem[\protect\citeauthoryear{{Koskinen}}{{Koskinen}}{2004}]{Ko04}
{Koskinen}, J. (2004).
\newblock {\em Essays on {B}ayesian Inference for Social Networks}.
\newblock Ph.\ D. thesis, Stockholm University, Dept.\ of Statistics, Sweden.

\bibitem[\protect\citeauthoryear{Koskinen, Robins, and Pattison}{Koskinen
  et~al.}{2010}]{KoRoPa09}
Koskinen, J.~H., G.~L. Robins, and P.~E. Pattison (2010).
\newblock Analysing exponential random graph (p-star) models with missing data
  using {B}ayesian data augmentation.
\newblock {\em Statistical Methodology\/}~{\em 7}, 366--384.

\bibitem[\protect\citeauthoryear{Lauritzen}{Lauritzen}{1996}]{La96}
Lauritzen, S. (1996).
\newblock {\em Graphical Models}.
\newblock Oxford, UK: Oxford University Press.

\bibitem[\protect\citeauthoryear{Lederer}{Lederer}{2021}]{Lederer21}
Lederer, J. (2021).
\newblock {\em Fundamentals of High-Dimensional Statistics---With Exercises and
  R Labs}.
\newblock Springer Texts in Statistics. Heidelberg: Springer.

\bibitem[\protect\citeauthoryear{Liang}{Liang}{2010}]{liang2010double}
Liang, F. (2010).
\newblock A double {M}etropolis--{H}astings sampler for spatial models with
  intractable normalizing constants.
\newblock {\em Journal of Statistical Computation and Simulation\/}~{\em
  80\/}(9), 1007--1022.

\bibitem[\protect\citeauthoryear{Liang and Jin}{Liang and
  Jin}{2013}]{liang2013monte}
Liang, F. and I.~H. Jin (2013).
\newblock A {M}onte {C}arlo {M}etropolis-{H}astings algorithm for sampling from
  distributions with intractable normalizing constants.
\newblock {\em Neural Computation\/}~{\em 25\/}(8), 2199--2234.

\bibitem[\protect\citeauthoryear{Liang, Jin, Song, and Liu}{Liang
  et~al.}{2016}]{liang2015adaptive}
Liang, F., I.~H. Jin, Q.~Song, and J.~S. Liu (2016).
\newblock An adaptive exchange algorithm for sampling from distributions with
  intractable normalizing constants.
\newblock {\em Journal of the American Statistical Association\/}~{\em 111},
  377--393.

\bibitem[\protect\citeauthoryear{Lusher, Koskinen, and Robins}{Lusher
  et~al.}{2013}]{ergm.book}
Lusher, D., J.~Koskinen, and G.~Robins (2013).
\newblock {\em Exponential Random Graph Models for Social Networks}.
\newblock Cambridge, UK: Cambridge University Press.

\bibitem[\protect\citeauthoryear{Lyne, Girolami, Atchade, Strathmann, and
  Simpson}{Lyne et~al.}{2015}]{Lyne15}
Lyne, A., M.~Girolami, Y.~Atchade, H.~Strathmann, and D.~Simpson (2015).
\newblock On {R}ussian roulette estimates for {B}ayesian inference with
  doubly-intractable likelihoods.
\newblock {\em Statistical Science\/}~{\em 30}, 1--26.

\bibitem[\protect\citeauthoryear{Maathuis, Drton, Lauritzen, and
  Wainwright}{Maathuis et~al.}{2019}]{graphical.models}
Maathuis, M., M.~Drton, S.~Lauritzen, and M.~Wainwright (2019).
\newblock {\em Handbook of Graphical Models}.
\newblock Boca Raton, Florida: CRC Press.

\bibitem[\protect\citeauthoryear{Marin, Pudlo, Robert, and Ryder}{Marin
  et~al.}{2012}]{marin2012approximate}
Marin, J.~M., P.~Pudlo, C.~P. Robert, and R.~J. Ryder (2012).
\newblock Approximate {B}ayesian computational methods.
\newblock {\em Statistics and Computing\/}~{\em 22\/}(6), 1167--1180.

\bibitem[\protect\citeauthoryear{Marjoram, Molitor, Plagnol, and
  Tavare}{Marjoram et~al.}{2003}]{Maetal03}
Marjoram, P., J.~Molitor, V.~Plagnol, and S.~Tavare (2003).
\newblock Markov chain {M}onte {C}arlo without likelihoods.
\newblock {\em Proceedings of the National Academices of Science, USA\/}~{\em
  100}, 15324--15328.

\bibitem[\protect\citeauthoryear{Marsman, Borsboom, Kruis, Epskamp, van Bork,
  Waldorp, van~der Maas, and Maris}{Marsman et~al.}{2018}]{Marsman:2018}
Marsman, M., D.~Borsboom, J.~Kruis, S.~Epskamp, R.~van Bork, L.~J. Waldorp,
  H.~L.~J. van~der Maas, and G.~K.~J. Maris (2018).
\newblock An introduction to network psychometrics: Relating ising network
  models to item response theory models.
\newblock {\em Multivariate Behavioral Research\/}~{\em 53}, 15--35.

\bibitem[\protect\citeauthoryear{Meinshausen and B\"uhlmann}{Meinshausen and
  B\"uhlmann}{2006}]{MeBu06}
Meinshausen, N. and P.~B\"uhlmann (2006).
\newblock High-dimensional graphs and variable selection with the {LASSO}.
\newblock {\em The Annals of Statistics\/}~{\em 34}, 1436--1462.

\bibitem[\protect\citeauthoryear{M{\o}ller, Pettitt, Reeves, and
  Berthelsen}{M{\o}ller et~al.}{2006}]{moller2006efficient}
M{\o}ller, J., A.~N. Pettitt, R.~Reeves, and K.~K. Berthelsen (2006).
\newblock An efficient {M}arkov chain {M}onte {C}arlo method for distributions
  with intractable normalising constants.
\newblock {\em Biometrika\/}~{\em 93\/}(2), 451--458.

\bibitem[\protect\citeauthoryear{Murray, Ghahramani, and MacKay}{Murray
  et~al.}{2006}]{murray2006}
Murray, I., Z.~Ghahramani, and D.~J.~C. MacKay (2006).
\newblock {MCMC} for doubly-intractable distributions.
\newblock In {\em Proceedings of the 22nd Annual Conference on Uncertainty in
  Artificial Intelligence}, pp.\  359--366. Corvallis: AUAI Press.

\bibitem[\protect\citeauthoryear{O'Hara, Sillanp{\"a}{\"a}, et~al.}{O'Hara
  et~al.}{2009}]{o2009review}
O'Hara, R.~B., M.~J. Sillanp{\"a}{\"a}, et~al. (2009).
\newblock A review of {B}ayesian variable selection methods: what, how and
  which.
\newblock {\em Bayesian analysis\/}~{\em 4\/}(1), 85--117.

\bibitem[\protect\citeauthoryear{Park and Haran}{Park and
  Haran}{2018}]{park2018bayesian}
Park, J. and M.~Haran (2018).
\newblock Bayesian inference in the presence of intractable normalizing
  functions.
\newblock {\em Journal of the American Statistical Association\/}~{\em
  113\/}(523), 1372--1390.

\bibitem[\protect\citeauthoryear{Park and Haran}{Park and
  Haran}{2020}]{park2019function}
Park, J. and M.~Haran (2020).
\newblock A function emulation approach for doubly intractable distributions.
\newblock {\em Journal of Computational and Graphical Statistics\/}~{\em
  29\/}(1), 66--77.

\bibitem[\protect\citeauthoryear{Park and Casella}{Park and
  Casella}{2008}]{park2008bayesian}
Park, T. and G.~Casella (2008).
\newblock The {B}ayesian lasso.
\newblock {\em Journal of the American Statistical Association\/}~{\em
  103\/}(482), 681--686.

\bibitem[\protect\citeauthoryear{Piaget}{Piaget}{1971}]{piaget:71}
Piaget, J. (1971).
\newblock {\em Biology and knowledge}.
\newblock University of Chicago Press: Chicago.

\bibitem[\protect\citeauthoryear{Pritchard, Seielstad, Perez-Lezaun, and
  Feldman}{Pritchard et~al.}{1999}]{Preetal99}
Pritchard, J., M.~T. Seielstad, A.~Perez-Lezaun, and M.~W. Feldman (1999).
\newblock Population growth of human {Y} chromosomes: {A} study of {Y}
  chromosome microsatellites.
\newblock {\em Molecular Biology and Evolution\/}~{\em 16}, 1791--1798.

\bibitem[\protect\citeauthoryear{Propp and Wilson}{Propp and
  Wilson}{1996}]{propp1996exact}
Propp, J.~G. and D.~B. Wilson (1996).
\newblock Exact sampling with coupled {M}arkov chains and applications to
  statistical mechanics.
\newblock {\em Random structures and Algorithms\/}~{\em 9\/}(1-2), 223--252.

\bibitem[\protect\citeauthoryear{Rand}{Rand}{1971}]{Ra71}
Rand, W.~M. (1971).
\newblock Objective criteria for the evaluation of clustering methods.
\newblock {\em Journal of the American Statistical Association\/}~{\em 66},
  846--850.

\bibitem[\protect\citeauthoryear{Ravikumar, Wainwright, and Lafferty}{Ravikumar
  et~al.}{2010}]{RaWaLa10}
Ravikumar, P., M.~J. Wainwright, and J.~Lafferty (2010).
\newblock High-dimensional {I}sing model selection using $\ell_1$-regularized
  logistic regression.
\newblock {\em The Annals of Statistics\/}~{\em 38}, 1287--1319.

\bibitem[\protect\citeauthoryear{Robert, Cornuet, Marin, and Pillai}{Robert
  et~al.}{2011}]{Robert15112}
Robert, C.~P., J.~M. Cornuet, J.~M. Marin, and N.~S. Pillai (2011).
\newblock Lack of confidence in approximate {B}ayesian computation model
  choice.
\newblock {\em Proceedings of the National Academy of Sciences\/}~{\em
  108\/}(37), 15112--15117.

\bibitem[\protect\citeauthoryear{Robins, Pattison, Kalish, and Lusher}{Robins
  et~al.}{2007}]{robins2007introduction}
Robins, G., P.~Pattison, Y.~Kalish, and D.~Lusher (2007).
\newblock An introduction to exponential random graph (p*) models for social
  networks.
\newblock {\em Social Networks\/}~{\em 29\/}(2), 173--191.

\bibitem[\protect\citeauthoryear{{Schweinberger} and
  {Handcock}}{{Schweinberger} and {Handcock}}{2015}]{ScHa13}
{Schweinberger}, M. and M.~S. {Handcock} (2015).
\newblock Local dependence in random graph models: characterization, properties
  and statistical inference.
\newblock {\em Journal of the Royal Statistical Society, Series B\/}~{\em 77},
  647--676.

\bibitem[\protect\citeauthoryear{Schweinberger, Krivitsky, Butts, and
  Stewart}{Schweinberger et~al.}{2020}]{ScKrBu17}
Schweinberger, M., P.~N. Krivitsky, C.~T. Butts, and J.~R. Stewart (2020).
\newblock Exponential-family models of random graphs: Inference in finite,
  super, and infinite population scenarios.
\newblock {\em Statistical Science\/}~{\em 35}, 627--662.

\bibitem[\protect\citeauthoryear{Shao}{Shao}{2003}]{shao}
Shao, J. (2003).
\newblock {\em Mathematical statistics\/} (2 ed.).
\newblock New York: Springer.

\bibitem[\protect\citeauthoryear{Sisson, Fan, and Tanaka}{Sisson
  et~al.}{2007}]{Sisson1760}
Sisson, S.~A., Y.~Fan, and M.~M. Tanaka (2007).
\newblock Sequential {M}onte {C}arlo without likelihoods.
\newblock {\em Proceedings of the National Academy of Sciences\/}~{\em 104},
  1760--1765.

\bibitem[\protect\citeauthoryear{Social and {Community Planning
  Research}}{Social and {Community Planning Research}}{1987}]{social1987}
Social and {Community Planning Research} (1987).
\newblock {\em British Social Attitude, the 1987 Report}.
\newblock Gower Publishing.

\bibitem[\protect\citeauthoryear{Spiel and Gluck}{Spiel and
  Gluck}{2008}]{spiel:08}
Spiel, C. and J.~Gluck (2008).
\newblock A model based test of competence profile and competence level in
  deductive reasoning.
\newblock In J.~Hartig, E.~Klieme, and D.~Leutner (Eds.), {\em Assessment of
  competencies in educational contexts: {State} of the art and future
  prospects}, pp.\  41--60. Gottingen: Hogrefe.

\bibitem[\protect\citeauthoryear{Spiel, Gluck, and Gossler}{Spiel
  et~al.}{2001}]{spiel:01}
Spiel, C., J.~Gluck, and H.~Gossler (2001).
\newblock Stability and change of unidimensionality: {The} sample case of
  deductive reasoning.
\newblock {\em Journal of Adolescent Research\/}~{\em 16}, 150--168.

\bibitem[\protect\citeauthoryear{Stewart, Schweinberger, Bojanowski, and
  Morris}{Stewart et~al.}{2019}]{StScBoMo18}
Stewart, J.~R., M.~Schweinberger, M.~Bojanowski, and M.~Morris (2019).
\newblock Multilevel network data facilitate statistical inference for curved
  {ERGM}s with geometrically weighted terms.
\newblock {\em Social Networks\/}~{\em 59}, 98--119.

\bibitem[\protect\citeauthoryear{Stoehr, Benson, and Friel}{Stoehr
  et~al.}{2017}]{stoehr2017noisy}
Stoehr, J., A.~Benson, and N.~Friel (2017).
\newblock Noisy {H}amiltonian {M}onte {C}arlo for doubly-intractable
  distributions.
\newblock {\em arXiv preprint arXiv:1706.10096\/}.

\bibitem[\protect\citeauthoryear{Strauss}{Strauss}{1975}]{strauss1975model}
Strauss, D.~J. (1975).
\newblock A model for clustering.
\newblock {\em Biometrika\/}~{\em 62\/}(2), 467--475.

\bibitem[\protect\citeauthoryear{Sundberg}{Sundberg}{2019}]{Su19}
Sundberg, R. (2019).
\newblock {\em Statistical Modelling by Exponential Families}.
\newblock Cambridge, UK: Cambridge University Press.

\bibitem[\protect\citeauthoryear{Thiemichen, Friel, Caimo, and
  Kauermann}{Thiemichen et~al.}{2016}]{thiemichen2016bayesian}
Thiemichen, S., N.~Friel, A.~Caimo, and G.~Kauermann (2016).
\newblock Bayesian exponential random graph models with nodal random effects.
\newblock {\em Social Networks\/}~{\em 46}, 11--28.

\bibitem[\protect\citeauthoryear{Toni, Welch, Strelkowa, Ipsen, and
  Stumpf}{Toni et~al.}{2009}]{Toni09}
Toni, T., D.~Welch, N.~Strelkowa, A.~Ipsen, and M.~P.~H. Stumpf (2009).
\newblock Approximate {B}ayesian computation scheme for parameter inference and
  model selection in dynamical systems.
\newblock {\em Journal of the Royal Society Interface\/}~{\em 6}, 187--202.

\bibitem[\protect\citeauthoryear{{van Borkulo}, Epskamp, and Robitzsch}{{van
  Borkulo} et~al.}{2016}]{isingfit}
{van Borkulo}, C., S.~Epskamp, and A.~Robitzsch (2016).
\newblock {\em IsingFit: Fitting Ising Models Using the ELasso Method}.
\newblock R package version 0.3.1.

\bibitem[\protect\citeauthoryear{{van Borkulo}, Borsboom, Epskamp, Blanken,
  Boschloo, Schoevers, and Waldorp}{{van Borkulo} et~al.}{2014}]{nature14}
{van Borkulo}, C.~D., D.~Borsboom, S.~Epskamp, T.~F. Blanken, L.~Boschloo,
  R.~A. Schoevers, and L.~J. Waldorp (2014).
\newblock A new method for constructing networks from binary data.
\newblock {\em Scientific Reports\/}~{\em 4}.

\bibitem[\protect\citeauthoryear{Wainwright}{Wainwright}{2019}]{Wa19}
Wainwright, M.~J. (2019).
\newblock {\em High-Dimensional Statistics. A Non-Asymptotic Viewpoint}.
\newblock Cambridge, UK: Cambridge University Press.

\bibitem[\protect\citeauthoryear{Whittaker}{Whittaker}{2009}]{Wh09}
Whittaker, J. (2009).
\newblock {\em Graphical Models in Applied Multivariate Statistics}.
\newblock New York: John Wiley \& Sons.

\bibitem[\protect\citeauthoryear{Xue, Zou, and Cai}{Xue
  et~al.}{2012}]{Xuetal12}
Xue, L., H.~Zou, and T.~Cai (2012).
\newblock Nonconcave penalized composite conditional likelihood estimation of
  sparse {I}sing models.
\newblock {\em The Annals of Statistics\/}~{\em 40}, 1403--1429.

\bibitem[\protect\citeauthoryear{Yi and Xu}{Yi and Xu}{2008}]{yi2008bayesian}
Yi, N. and S.~Xu (2008).
\newblock Bayesian {LASSO} for quantitative trait loci mapping.
\newblock {\em Genetics\/}~{\em 179\/}(2), 1045--1055.

\bibitem[\protect\citeauthoryear{Yin and Butts}{Yin and Butts}{2020}]{YiBu20}
Yin, F. and C.~T. Butts (2020).
\newblock Kernel-based approximate {B}ayesian inference for {E}xponential
  {F}amily {R}andom {G}raph {M}odels.
\newblock arXiv:2004.08064.

\bibitem[\protect\citeauthoryear{Zhao and Yu}{Zhao and Yu}{2006}]{ZhYu06}
Zhao, P. and B.~Yu (2006).
\newblock On model selection consistency of the {L}asso.
\newblock {\em Journal of Machine Learning Research\/}~{\em 7}, 2541--2563.

\end{thebibliography}

\end{document}